\documentclass[sigconf]{acmart}

\AtBeginDocument{%
\providecommand\BibTeX{{%
\normalfont B\kern-0.5em{\scshape i\kern-0.25em b}\kern-0.8em\TeX}}}

\usepackage{tabularx}
\usepackage{enumitem}
\usepackage{xspace}
\usepackage{appendix}
\usepackage{multirow}          
\usepackage{booktabs} 
\usepackage{csquotes}
\usepackage{cleveref}
\usepackage{listings}
\usepackage{xcolor}
\usepackage{float}

\definecolor{codebackground}{rgb}{0.95,0.95,0.95}
\definecolor{codekeyword}{rgb}{0.13,0.29,0.53}
\definecolor{codestring}{rgb}{0.63,0.125,0.94}
\definecolor{codecomment}{rgb}{0.25,0.5,0.37}

\lstdefinestyle{promptstyle}{
    backgroundcolor=\color{codebackground},
    basicstyle=\small\ttfamily,
    breakatwhitespace=false,
    breaklines=true,
    captionpos=b,
    commentstyle=\color{codecomment},
    frame=single,
    keepspaces=true,
    keywordstyle=\color{codekeyword},
    showspaces=false,
    showstringspaces=false,
    showtabs=false,
    stepnumber=1,
    stringstyle=\color{codestring},
    tabsize=2,
    columns=flexible,
    resetmargins=true,
    framesep=3pt,
    framexleftmargin=3pt,
    framexrightmargin=3pt,
    xleftmargin=3pt,
    xrightmargin=3pt
}

\lstnewenvironment{prompt}
{\lstset{style=promptstyle}}
{}

\lstdefinestyle{jsonstyle}{
    backgroundcolor=\color{codebackground},
    basicstyle=\small\ttfamily,
    breakatwhitespace=false,
    breaklines=true,
    captionpos=b,
    commentstyle=\color{codecomment},
    frame=single,
    keepspaces=true,
    keywordstyle=\color{codekeyword},
    showspaces=false,
    showstringspaces=false,
    showtabs=false,
    stringstyle=\color{codestring},
    tabsize=2,
    morestring=[b]",
        columns=flexible,
    resetmargins=true,
    framesep=3pt,
    framexleftmargin=3pt,
    framexrightmargin=3pt,
    xleftmargin=3pt,
    xrightmargin=3pt
}

\lstnewenvironment{jsoncode}
{\lstset{style=jsonstyle}}
{}

\copyrightyear{2025}
\acmYear{2025}

\setcopyright{cc}
\setcctype{by}
\acmConference[ASSETS '25]{The 27th International ACM SIGACCESS Conference on Computers and Accessibility}{October 26--29, 2025}{Denver, CO, USA}
\acmBooktitle{The 27th International ACM SIGACCESS Conference on Computers and Accessibility (ASSETS '25), October 26--29, 2025, Denver, CO, USA}
\acmDOI{10.1145/3663547.3746380}
\acmISBN{979-8-4007-0676-9/2025/10}

\begin{document}
\newcommand\lucy[1]{{\color{blue}\{\textit{#1}\}$_{lucy}$}}

\newcommand\todoit[1]{{\color{red}\{TODO: \textit{#1}\}}}
\newcommand\todo{{\color{red}{TODO}}\xspace}
\newcommand\todocite{{\color{red}{CITE}}\xspace}

\newcommand\uwname{{our institution\xspace}}

\newcolumntype{L}[1]{>{\raggedright\let\newline\\\arraybackslash\hspace{0pt}}p{#1}}
\newcolumntype{M}[1]{>{\raggedright\let\newline\\\arraybackslash\hspace{0pt}}m{#1}}
\newcolumntype{C}[1]{>{\centering\arraybackslash\hspace{0pt}}p{#1}}

\newcommand{\rulesep}{\unskip\ \textcolor{gray}{\vrule}\ }

\definecolor{darkgreen}{rgb}{0.0, 0.4, 0.13}
\title{Benchmarking PDF Accessibility Evaluation} 
\subtitle{A Dataset and Framework for Assessing Automated and LLM-Based Approaches for Accessibility Testing}

\author{Anukriti Kumar}
\email{anukumar@uw.edu}
\orcid{}
\affiliation{%
  \institution{University of Washington}
  \city{Seattle}
  \state{WA}
  \postcode{98103}
  \country{USA}
}

\author{Tanushree Padath}
\email{tpadath@uw.edu}
\orcid{}
\affiliation{%
  \institution{University of Washington}
  \city{Seattle}
  \state{WA}
  \postcode{98103}
  \country{USA}
}

\author{Lucy Lu Wang}
\email{lucylw@uw.edu}
\orcid{0000-0001-8752-6635}
\affiliation{%
  \institution{University of Washington}
  \city{Seattle}
  \state{WA}
  \postcode{98103}
  \country{USA}
}

\renewcommand{\shortauthors}{Anukriti Kumar and Lucy Lu Wang}

\begin{abstract}
PDFs remain the dominant format for scholarly communication, despite significant accessibility challenges for blind and low-vision users.
While various tools attempt to evaluate PDF accessibility, there is no standardized methodology to evaluate how different accessibility assessment approaches perform. Our work addresses this critical gap by introducing a novel benchmark dataset of scholarly PDFs with expert-validated accessibility annotations across seven criteria (alternative text quality, logical reading order, semantic tagging, table structure, functional hyperlinks, color contrast, and font readability), and a four-category evaluation framework with standardized labels (Passed, Failed, Not Present, Cannot Tell) to systematically assess accessibility evaluation approaches. Using our evaluation framework, we explore whether large language models (LLMs) are capable of supporting automated accessibility evaluation. We benchmark five LLMs, which demonstrate varying capabilities in correctly assessing different accessibility criteria, with GPT-4-Turbo achieving the highest overall accuracy (0.85).
However, all models struggled in correctly categorizing documents with Not Present and Cannot Tell accessibility labels, particularly for alt text quality assessment. Our qualitative comparison with standard automated checkers reveals complementary strengths: rule-based tools excel at technical verification, while LLMs better evaluate semantic appropriateness and contextual relevance. Based on our findings, we propose a hybrid approach that would combine automated checkers, LLM evaluation, and human assessment as a future strategy for PDF accessibility evaluation.\footnote{The dataset and code for running experiments can be found at \url{https://github.com/Anukriti12/PDF-Accessibility-Benchmark}}
\end{abstract}



\received{16 April 2025}

\maketitle

\section{Introduction}
\label{sec:introduction}


Scholarly literature disseminated as PDF documents systematically excludes readers with disabilities—particularly blind and low-vision (BLV) users who rely on screen readers and keyboard navigation—creating substantial barriers to education, research participation, and professional advancement. Recent studies reveal the scope of this problem, finding that less than 3.2\% of scholarly PDFs published between 2014 and 2023 meet the key tested accessibility criteria, and over 75\% fail to satisfy even a single requirement \cite{anu_uncovering}. These findings align with earlier works that found less than 30\% compliance across accessibility criteria in major computing conferences \cite{Brady2015CreatingAP, Lazar2017MakingTF}, with only 15-30\% compliance rates in disability studies journals \citet{Nganji2015ThePD}. 

\begin{figure*}[t!]
  \centering
    \includegraphics[width=0.76\linewidth]{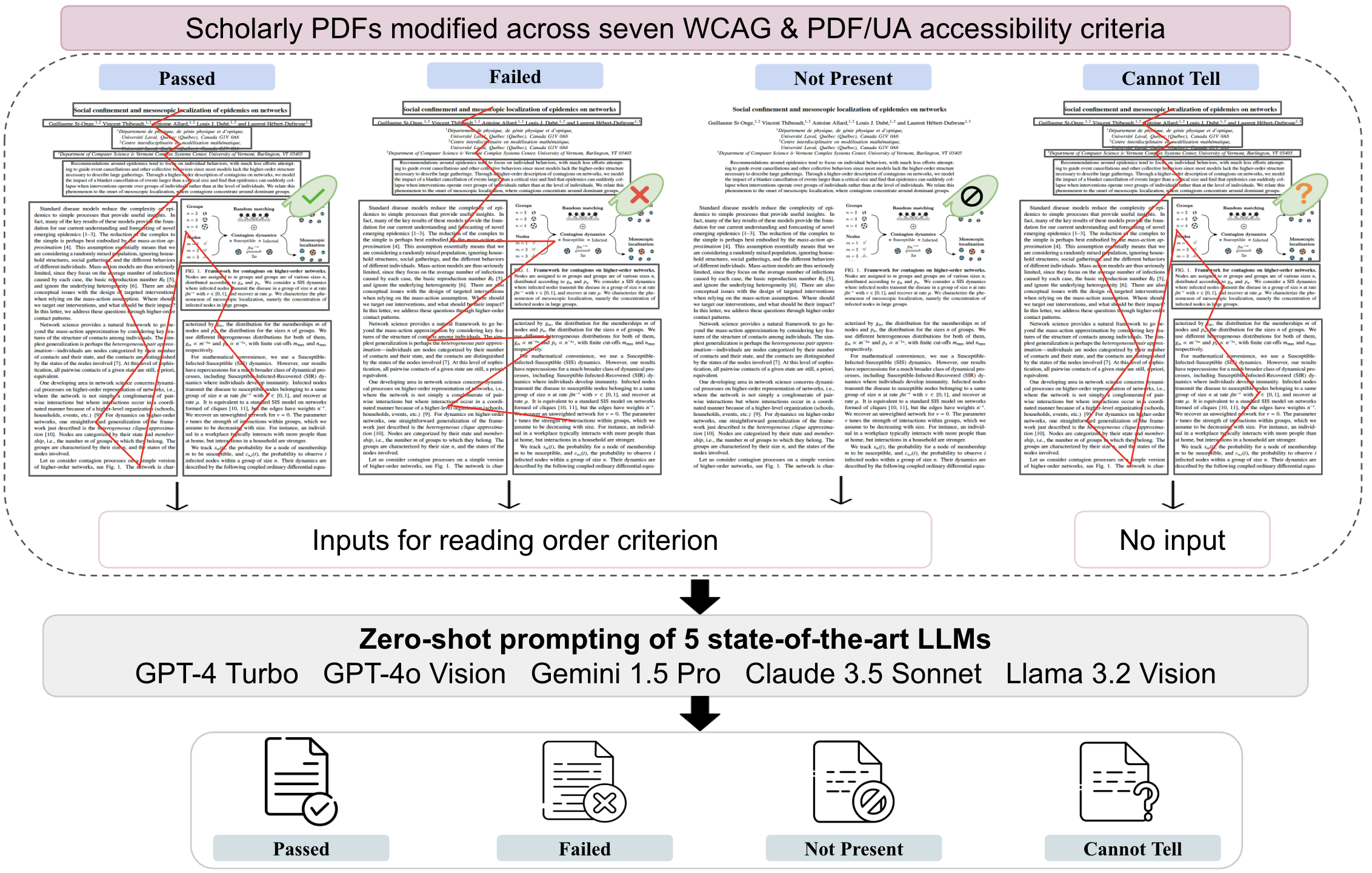}
  \caption{Our framework for PDF Accessibility Evaluation using our curated benchmark and several Large Language Models. Example inputs corresponding to the four label classes (Passed, Failed, Not Present, and Cannot Tell) are shown for the Reading Order criterion. Red lines indicate the reading order sequence on a page.}
  
  \label{framework_diagram}
  \Description{
  Workflow diagram showing the evaluation framework for assessing PDF accessibility using large language models. The top section displays four modified versions of the same scholarly PDF page, each representing a different accessibility label: Passed, Failed, Not Present, and Cannot Tell. Red arrows trace the logical reading order through each version. The “Passed” and “Failed” documents contain highlighted tag trees and visual flow indicators; the “Not Present” document lacks tags; the “Cannot Tell” document omits input features like the tag tree and reading order information entirely. Corresponding inputs for reading order criterion are extracted from all the documents, except the document with "Cannot Tell" label. These inputs are fed into five large language models—GPT-4 Turbo, GPT-4o Vision, Gemini 1.5 Pro, Claude 3.5 Sonnet, and LLaMA 3.2 Vision—using zero-shot prompting. At the bottom, four icons depict the output labels produced by the models: Passed, Failed, Not Present, and Cannot Tell.
  }
\end{figure*}



While established guidelines like Web Content Accessibility Guidelines (WCAG) 2.2 \cite{Caldwell2008WebCA} and PDF/Universal Accessibility \cite{PDF_UA1} provide comprehensive specifications for accessibility compliance, implementation remains severely limited. This gap persists despite legal frameworks such as the Americans with Disabilities Act (ADA) \cite{ADA1990}, Section 508 in the United States \cite{Section508.gov}, and the European Accessibility Act \cite{European_accessibility_act} that makes accessible content a civil right. This limited adherence largely stems from a combination of factors: limited awareness and training among content creators \cite{Lazar2017MakingTF}, inadequate authoring tools \cite{Angerbauer2022colorvision}, and the inherent complexity of the PDF format itself \cite{Bigham2016AnUT}, which was originally designed for visual faithfulness rather than accessibility.

A critical challenge in addressing this accessibility crisis is the lack of standardized evaluation methods and benchmarks for PDF accessibility. Current approaches rely predominantly on rule-based automated checkers like Adobe Acrobat Pro, PAC 2024, and axesPDF. While these tools effectively identify basic compliance issues such as the presence of tags or alt attributes \cite{Alsaeedi_2020}, they cannot be used to evaluate semantic, contextual, and user-centered accessibility \cite{anu_uncovering}. Research indicates that automated checkers can only detect approximately 25–30\% of issues \cite{Gratzer_2021}, particularly struggling with nuanced assessments such as the appropriateness of alternative text for images \cite{Yadavali_2024}. As a result, manual testing has become essential for identifying issues that automated tools cannot detect. For example, manual testing can reveal problems related to user experience, such as whether a screen reader can navigate a page correctly or whether a user with a motor impairment can interact with interactive elements \cite{10.1145/3638067.3638101}. Manual testing also allows for the identification of issues that require human judgment, such as the clarity of written content or the logical structure of a page \cite{Yadavali_2024, Sun_Vu_Strybel_2018}.
However, manual testing can be time-consuming and resource-intensive, making it difficult to scale—particularly in scholarly publishing, where thousands of documents are produced annually \cite{arxiv_report}. 

Recent advances in large language models (LLMs) offer potential new opportunities to tackle these challenges. Researchers have demonstrated LLMs' capabilities across various accessibility domains through specialized applications in generating alternative text for images \cite{FigurA11y}, identifying code-level accessibility issues \cite{CodeA11y}, and using LLMs to remediate PDF forms for screen reader compatibility \cite{FormA11y}. Despite these advances, most existing work still focuses on web accessibility rather than PDF documents, which present unique challenges due to their fixed layout and complex structure. For instance, LLMs like ChatGPT have been used to automatically remediate web accessibility issues, ensuring compliance with WCAG guidelines and demonstrating effectiveness in comparison to manual testing methods \cite{Othman}. Additionally, unlike web content that has WCAG accessibility conformance testing rules (WCAG-ACT rules) and associated datasets \cite{act-tests} for benchmarking, there is a lack of benchmark datasets and structured evaluation protocols to systematically assess whether any tool or LLMs can enhance PDF accessibility evaluation.

 

Our work aims to address these gaps by developing a novel benchmark dataset of scholarly PDFs with expert-validated accessibility annotations across seven criteria derived from the WCAG 2.2 and PDF/UA standards: alternative text quality, logical reading order, semantic tagging, table structure, functional hyperlinks, color contrast, and font readability. We further investigate whether LLMs can automate the evaluation of these criteria that traditionally require human judgment. For each criterion, we evaluate LLM performance across four accessibility labels, detailed in the W3C report \cite{w3WCAGEMReportTool}: Passed (meets all requirements), Failed (violates requirements), Not Present (element absent), and Cannot Tell (insufficient information to evaluate). Our primary contributions are summarized below:

\begin{itemize}[topsep=2pt,itemsep=2pt, leftmargin=10pt]
    \item We create a benchmark dataset of scholarly PDFs with expert-validated accessibility annotations across seven criteria aligned with WCAG and PDF/UA standards: alternative text quality, logical reading order, semantic tagging, table structure, functional hyperlink, color contrast, and font readability. Each document is labeled within a four-category evaluation framework (Passed, Failed, Not Present, Cannot Tell), recommended by W3C \cite{w3WCAGEMReportTool}. This represents the first publicly available dataset specifically designed for evaluating accessibility assessment approaches on PDF documents.

    \item We develop and evaluate a zero-shot LLM framework for PDF accessibility assessment using our benchmark dataset (Figure~\ref{framework_diagram}). By comparing performance across five models---GPT-4-Turbo \cite{gpt4}, GPT-4o-Vision \cite{gpt4o}, Claude-3.5 sonnet \cite{claude-3.5}, Gemini 1.5 \cite{gemini}, and Llama-3.2 \cite{llama}---across the seven criteria mentioned above, we demonstrate where current LLMs excel (structure evaluation, font compliance) and where they struggle (alternative text assessment, recognizing information gaps). Unlike previous work that treats accessibility as a binary property, our approach captures the nuanced, multi-faceted nature of accessibility compliance. 
    

    \item We provide a qualitative comparison between LLM-based assessment and traditional automated checkers. This analysis reveals complementary strengths: while rule-based tools excel at verifying technical compliance, LLMs better evaluate semantic appropriateness and contextual relevance. For instance, where automated checkers merely verify alt text exists, LLMs can identify when descriptions like "Figure" fail to convey the image's meaning. Similarly, LLMs can assess if reading order logically follows content flow, a capability beyond current automated tools. These complementary strengths suggest potential for hybrid systems that combine rule-based verification with contextual reasoning.

\end{itemize}

\noindent By establishing this benchmark dataset and evaluation framework, our work enables systematic comparison of accessibility evaluation approaches, identifies opportunities for improving automated assessment, and provides a foundation for developing more effective tools to address the PDF accessibility crisis.

The remainder of this paper proceeds as follows. Section \ref{sec:related_work} discusses related work in PDF accessibility evaluation, automated checkers, and LLMs in PDF accessibility. Sections \ref{sec:dataset} and \ref{sec:experiments} present our dataset creation methodology and experimental approach for LLM-based evaluation respectively. Section \ref{sec:results} reports experimental results, analyzing LLM performance across criteria and accessibility labels. Section \ref{sec:comparison} provide a qualitative comparison with existing tools. Sections \ref{sec:discussion} and \ref{sec:limitations} discuss implications, limitations, and future directions. Section \ref{sec:conclusion} concludes with reflections on integrating LLMs into accessibility evaluation workflows.

\section{Background and Related work}
\label{sec:related_work}

This section examines prior research in PDF accessibility evaluation, automated checkers, and the emerging use of LLMs for accessibility tasks. We identify the gap in standardized evaluation methods for PDF accessibility that our work addresses.

\subsection{PDF Accessibility Evaluation}
PDF accessibility has long been a critical concern in digital content delivery, particularly within academic publishing. A large majority (74.9\%) fail to meet any accessibility criteria at all \cite{anu_uncovering}, corroborating earlier studies reporting only 15–30\% compliance in major journals and conference proceedings \cite{Nganji2015ThePD}. This presents a significant challenge given the volume of academic literature published annually—arXiv alone hosts over 2 million scholarly articles with approximately 15,000 new submissions monthly \cite{arxiv_report}, creating substantial barriers for readers with disabilities. 

Accessibility evaluation tools play a crucial role in ensuring that digital products are usable by everyone, including individuals with disabilities. These tools can check for violations of guidelines such as the Web Content Accessibility Guidelines (WCAG), PDF/UA and provide authors with actionable insights. For instance, tools like Adobe Acrobat Pro, PAC 3/2024, CommonLook, and axesPDF have been shown to effectively employ rule-based logic to detect technical violations of accessibility standards, such as missing alt text for images and improper heading structures \cite {Alsaeedi_2020, anu_uncovering}.
However, they demonstrate significant limitations in comprehensiveness and accuracy.

Early comparative studies by \citet{Brajnik_2004} proposed methods to assess tool correctness, specificity, and completeness. Subsequent research demonstrated that automated tools typically detect only 25–30\% of web accessibility barriers \cite{Vigo, key}. These tools perform adequately for detecting structural issues like missing tags or unlabeled form fields but frequently fail to identify semantic problems such as misleading alternative text, improper reading order, or incorrect tag semantics \cite{anu_uncovering}. Additionally, automated checkers often differ in their interpretation of standards, employ inconsistent terminology, and lack transparency in their decision-making processes \cite{tools}. This variability complicates comparative analysis and undermines confidence in automated assessment, particularly for complex documents.

Recognizing the limitations of automated tools, accessibility experts advocate for manual evaluation as the gold standard, particularly for assessing semantic accuracy and user experience \cite{anu_uncovering, Vigo}. Additionally, it allows experts to evaluate how screen readers actually interact with content and identify issues that require human judgment, such as the clarity of written content, the logical structure of a page, or the contextual appropriateness of alternative text \cite{Vigo}. However, manual testing presents significant scalability challenges—it requires specialized expertise, is time-consuming, and becomes prohibitively expensive when applied to large document collections (\cite{bai, arxiv_report}. 

To address this tension between thoroughness and scalability, researchers have proposed hybrid workflows that combine automated detection with human verification \cite{lopez, Shah_Garg}. These approaches leverage automated tools to handle routine checks while reserving human judgment for more nuanced evaluation tasks. For instance, \cite{lopez} demonstrated how AI-based pre-screening can reduce the time and cost associated with manual accessibility evaluations for certain accessibility criteria. Meanwhile, industry initiatives like the PDF Association's Matterhorn Protocol \cite{PDF_Association} have established frameworks for categorizing accessibility issues by whether they require automated or human verification, further acknowledging the complementary nature of these approaches. The W3C also provides detailed guidance through Techniques for WCAG 2.0 for PDF documents \cite{W3C}, yet adoption in practice remains limited.


\subsection{AI for PDF Accessibility}
AI and large language models (LLMs) are increasingly being utilized to enhance PDF accessibility for users with disabilities. Several AI-driven tools have been developed to automate specific remediation tasks, such as generating alternative text for images \cite{FigurA11y} and suggesting more descriptive replacements to improve alt text \cite{SaGol}. In addition, AI-powered semantic frameworks can identify issues often overlooked by traditional checkers, including color contrast violations and improper heading hierarchies \cite{Yadavali_2024}. While these approaches show promise for targeted tasks, they often struggle to generalize across document formats and diverse accessibility criteria \cite{suh2025humanllmcomparativestudy}.

The emergence of advanced LLMs has opened new possibilities for automated accessibility evaluation. Unlike traditional rule-based systems, LLMs possess capabilities in natural language understanding, semantic reasoning, and multimodal processing—skills that align well with the nuanced demands of accessibility assessment. Recent studies have explored these capabilities primarily in web and mobile contexts. For example, \citet{suh2025humanllmcomparativestudy} compared LLM-generated and human-authored accessible code, showing that models effectively implemented basic accessibility principles, such as alternative text and color contrast, but struggled with more complex ARIA implementations. \citet{huang2024accesspromptengineeringautomated} applied LLMs to detect WCAG violations in web markup, while \citet{mobilea11y} evaluated their ability to provide code suggestions to fix mobile accessibility issues.

LLMs have also shown promising results on specific accessibility evaluation tasks, especially to detect structural accessibility violations, such as incorrect heading levels and improper tag nesting, finding strong alignment between model predictions and expert judgments \cite{duarte}. The AXNav system \cite{taeb_axnav} employed LLM-based planners to execute accessibility tests from natural language instructions, illustrating how such models can facilitate structured testing workflows. Various commercial solutions including Acquia Optimize \cite{Acquia} have begun incorporating AI-driven tools for web accessibility compliance, suggesting an industry shift toward other evaluation methods.

Despite these advances, LLM-based accessibility research has largely focused on web and mobile interfaces, with limited attention to PDFs. This may be partly due to the presence of well-established benchmarking datasets for web content—such as the WCAG 2.0 Test Samples Repository \cite{wcag-test-samples}, and the Benchmark for Evaluation Tools \cite{1234010} which have enabled rigorous evaluation of tool performance and standards compliance. These resources provide controlled sets of accessible and non-accessible web pages, supporting both automated tool benchmarking and academic research. However, no comparable publicly available benchmark exists for systematically evaluating PDF documents against WCAG 2.2 or PDF/UA criteria. This absence hinders reproducibility, limits tool validation, and slows the development of intelligent PDF accessibility checkers. Furthermore, the effectiveness of LLMs in handling the multimodal nature of PDF accessibility, that is, simultaneously reasoning over document images, tag trees, and structural metadata—remains underexplored. While foundational work by \citet{FormA11y} on form remediation and preliminary explorations by \citet{tools} on PDF assessment tools suggest potential, systematic evaluation across comprehensive accessibility criteria has not been conducted.

In light of these gaps, we construct the first expert-validated benchmark dataset spanning seven core PDF accessibility criteria and evaluate the capabilities of five leading LLMs—GPT4-Turbo, GPT-4o-vision, Claude-3.5, Gemini 1.5, and Llama-3.2—using a zero-shot prompting approach. Our evaluation labels each document as Passed, Failed, Not Present, or Cannot Tell, aligned with expert annotations. We also provide a qualitative comparison of LLM-based and traditional automated evaluations, identifying strengths, weaknesses, and opportunities for hybrid approaches that combine the technical precision of rule-based systems with the contextual reasoning strengths of LLMs.

\section{Benchmark Construction}
\label{sec:dataset}

This section details our methodology for creating a comprehensive benchmark dataset spanning seven accessibility criteria. Our approach focused on developing an evaluation framework that measures 
usability issues that impact people with disabilities, and which are not well assessed in current evaluation of technical compliance. We describe our criteria selection process, document modification approach, and relevant inputs extraction from these documents.

\subsection{Accessibility Criteria Selection}
Our study focuses on seven key accessibility criteria derived from WCAG 2.2 and PDF/UA standards, selected based on these considerations:
\begin{itemize}[topsep=2pt,itemsep=2pt, leftmargin=10pt]
\item They address areas where existing automated tools fail and require manual evaluation according to prior research \cite{anu_uncovering},
\item They represent diverse aspects of document accessibility with particular importance for screen reader users.
\end{itemize}

While WCAG was originally designed for web content, many of its principles apply to PDF documents and are cross-referenced in the PDF/UA standard. However, applying these guidelines to PDFs often involves subjective interpretation, particularly because PDFs present unique structural and semantic challenges not encountered in HTML-based content. 

To guide our evaluation, we developed a detailed sub-criteria list for each main criterion, summarized in Table~\ref{tab:criteria-summary}, that capture the nuanced requirements of accessible PDFs. Full definitions and evaluation checklists for each criterion are provided in Appendix Tables~\ref{tab:alt_text_criteria}-~\ref{tab:font_criteria}. These sub-criteria were informed by both accessibility standards and best practices identified in real-world PDF evaluation scenarios. They served as a foundation for both prompt development and expert evaluation guidelines used in our study.

\begin{table*}[t]
\centering
\small
\begin{tabular}{L{30mm}L{60mm}L{40mm}}
\toprule
\textbf{Criterion} & \textbf{Sub-Criteria} & \textbf{Reference Standards} \\
\midrule
\textbf{Alternative Text Quality} & Descriptiveness, complexity handling, decorative treatment, clarity, conciseness, contextual appropriateness & WCAG 2.2 SC 1.1.1, PDF/UA, Best Practices  \\
\addlinespace
\textbf{Semantic Tagging} & Structural hierarchy, artifacts, heading nesting, list structure, semantics (proper tag usage) & WCAG 2.2 SC 1.3.1, WCAG 2.2 SC 1.1.1, WCAG 2.2 SC 2.4.2, PDF/UA, Best Practices \\
\addlinespace
\textbf{Logical Reading Order} & Visual flow alignment, contextual logic, element positioning, hierarchy representation, no disruptions, multi-column handling, inline elements, hidden content & WCAG 2.2 SC 1.3.2, WCAG 2.2 SC 2.4.3, WCAG 2.2 SC 1.3.1, PDF/UA, Best Practices \\
\addlinespace
\textbf{Table Structure} & Proper table tagging, header semantics, contextual clarity, appropriate usage, keyboard navigability, complex table handling & WCAG 2.2 SC 1.3.1, WCAG 2.2 SC 1.3.2, WCAG 2.2 SC 2.4.3, WCAG 2.2 SC 2.1.1, WCAG 2.2 SC 1.3.3, WCAG 2.2 SC 2.4.6, PDF/UA, Best Practices \\
\addlinespace
\textbf{Functional Hyperlinks} & Proper link tagging, link functionality, link descriptiveness, structure association, link perceivability & WCAG 2.2 SC 1.4.3, WCAG 2.2 SC 2.4.4, WCAG 2.2 SC 2.4.9, WCAG 2.2 SC 1.3.1, WCAG 2.2 SC 4.1.2, PDF/UA, Best Practices \\
\addlinespace
\textbf{Color Contrast} & Small text contrast, large text contrast, informative visual elements, decorative elements, text in images & WCAG 2.2 SC 1.4.3, WCAG 2.2 SC 1.4.6, WCAG 2.2 SC 1.4.11, Best Practices \\
\addlinespace
\textbf{Font Readability} & Font embedding, legibility, font size, semantic encoding, font changes, consistency & WCAG 2.2 SC 1.4.12, WCAG 2.2 SC 1.4.8, WCAG 2.2 SC 1.4.4, WCAG 2.2 SC 3.2.4, PDF/UA, Best Practices \\
\bottomrule
\end{tabular}
\caption{Accessibility criteria, sub-criteria, and associated guidelines}
\label{tab:criteria-summary}
\end{table*}

\subsection{Corpus Creation}
To support benchmarking, we constructed a corpus of scholarly PDFs, each systematically modified to reflect compliance or non-compliance with specific accessibility criteria. This section outlines our three-stage corpus creation process: (1) sampling and selecting source documents that exhibit the structural and content features relevant to each criterion, (2) modifying these documents to represent each possible accessibility label (Passed, Failed, Not Present, Cannot Tell) as defined in Table~\ref{tab:criteria-summary}, and (3) validating the modified documents and assigned labels through expert review, using WCAG 2.2 and PDF/UA guidelines as the compliance baseline.

\subsubsection{Paper Corpus and Sampling Strategy}
We built our dataset using a strategic sampling approach from a larger corpus of scholarly papers. We began with an initial corpus of 20,000 scholarly PDFs published between 2014 and 2023, compiled by~\citet{anu_uncovering} using metadata from OpenAlex and pre-screened with Adobe Acrobat's accessibility checker to provide baseline accessibility information. We then applied a three-stage sampling process: (1) filtered by accessibility scores, (2) manually screened for key features (e.g., figures, tables, headings), and (3) selected highly cited papers across seven academic fields. This way, we curated 35 representative PDFs (five per accessibility criterion) that would serve as base documents for our evaluation framework. Further, we employed manual edits using Adobe Acrobat Pro to perform controlled modifications that would produce documents with specific accessibility characteristics, as mentioned in Table~\ref{tab:document_modifications}. Given the substantial manual effort required to document specific accessibility requirements (like editing tag trees, authoring alternative text, adjusting structural elements), we intentionally limited the sample size and instead focused more on the precision of our modifications while maintaining a manageable scope for expert validation.

We chose documents with varied layout, organizational structure, and content type to ensure the dataset represented the complexity found in academic literature. Each document contained features relevant to the accessibility criterion it would be used to evaluate (e.g., figures or diagrams for alternative text assessment, multi-level tables with header rows and merged cells for table structure evaluation). We also ensured representation across seven academic fields: computer science, medicine, sociology, environmental science, physics, biology, and history, and chose highly-cited papers within each field to mirror the documents most likely to be viewed by readers. The resulting document set had high natural variation in content presentation styles and document complexity.

\subsubsection{Accessibility Labels}
Following W3C recommendations \cite{w3WCAGEMReportTool}, we structured our dataset around four distinct labels, which were applied independently to each criterion:
\begin{itemize}
    \item Passed (P): Documents fully complying with all sub-criteria for the specific accessibility requirement
    \item Failed (F): Documents violating one or more sub-criteria (table~\ref{tab:criteria-summary})
    \item Not Present (NP): Documents where the relevant element being evaluated is absent (e.g., the paper has no alt text when evaluating alt-text quality)
    \item Cannot Tell (CT): Documents where insufficient information is provided at evaluation time to make a determination (e.g., no alt text is provided to the checker when evaluating alt-text quality)
\end{itemize}

These labels align with established accessibility evaluation frameworks, including the W3C Evaluation Methodology (WCAG-EM) and several automated checking tools. They represent the full range of possible outcomes in accessibility evaluation, including edge cases where assessment is either inapplicable or impossible due to missing information.

\subsubsection{Document Modification}
For each criterion, we modified the five documents selected using Adobe Acrobat Pro to create instances representing the four evaluation accessibility labels (P, F, NP, CT). This resulted in 4 copies of each paper PDF corresponding to the 4 labels. However, for some criteria, namely logical reading order, font readability, and color contrast, the `Not Present' label was not feasible. These criteria are inherent properties of every PDF document; all documents exhibit some logical flow of reading, contain text with visible color contrast, and use fonts that can be assessed for size, encoding, and embedding. As such, it is always possible to evaluate these aspects, even if they fail or contain structural ambiguities. As a result, the final dataset consisted of 125 PDFs (5 base documents per criteria × 4 labels x 4 criteria + 5 base documents per criteria x 3 labels x 3 criteria). 

The modifications we made varied by criterion and target label; for example, when creating variants for the Semantic Tagging criterion, we made the following modifications:
\begin{itemize}
    \item Passed (P): We ensured proper heading hierarchy, list structures, and appropriate content tags
    \item Failed (F): We introduced specific errors like incorrectly nested headings (h1→h3→h2), missing list containers, or non-semantic span tags
    \item Not Present (NP): We removed all structural tags, converting the document to untagged content
    \item Cannot Tell (CT): We did not modify the document but at evaluation time, we provided the checkers only with the document's images (visual rendering) and withheld the tag tree information
\end{itemize}

This was repeated on different PDFs for each of our 7 criteria. Descriptions of modifications made for other criteria can be found in the Appendix~\ref{sec:document_modification_list}.

\subsubsection{Internal Validation} To ensure the validity of our dataset, all documents and their accessibility labels were independently reviewed by an accessibility specialist with over five years of experience in document accessibility auditing and WCAG compliance consulting. Each modified document underwent independent inspection using the VoiceOver screen reader, cross-validation of our accessibility labels, and thorough inspection of tag trees, reading order, alternative text structures, and other semantic markers. Finally, any disagreements were resolved through discussion and further modification of the PDFs. This validation process ensured that our ground truth labels accurately reflected professional judgment and adherence to WCAG and PDF/UA standards.

\subsection{Document Representations}
\label{sec:document_representationa}
For each document in our dataset, we prepared standardized input representations tailored to the specific accessibility criterion being evaluated. These inputs, summarized in Table~\ref{tab:input-data}, include:
\begin{itemize}
    \item Document metadata: title, field of study, author, and publication venue
    \item Page-level raster images: generated using the PyMuPDF library to simulate a visual rendering of each PDF
    \item Criterion-specific representations: extracted using a combination of third-party APIs, accessibility tools, and custom processing scripts. This includes tag structures, table tags, color hex code combinations, reading order, font metadata, alternative text, and screen reader transcripts.
    
\end{itemize}


For Cannot Tell cases, we deliberately withheld criterion-specific information necessary for evaluation, thus simulating scenarios when only incomplete data can be extracted from the document. 
This approach allowed us to assess not only a checker's ability to identify accessibility issues but also their capability to recognize when they lack sufficient information to make a determination. 


To identify and extract figures and tables, we used a layout detection pipeline built on a Faster R-CNN architecture, fine-tuned on the PubLayNet dataset \cite{publaynet}. This model analyzed each PDF page and returned region-wise labels ("Table" and "Figure") along with their bounding boxes, which we used to crop image segments directly from each page. To obtain alternative text, we applied the method proposed by \citet{Chintalapati_2022}, using Adobe Acrobat Pro’s PDF-to-HTML conversion utility. From the resulting markup, we parsed the alt attributes associated with <img> tags and matched them to detected figure and table elements in the original PDF. 

Font metadata, including font family, size, weight, and embedding status, was obtained from the Adobe PDF Extract API, which provides word-level font annotations in a structured JSON format. Additionally, we used a combination of content-aware pixel-level analysis and clustering techniques to identify dominant foreground and background color combinations, represented in hexadecimal codes. For hyperlink and navigational metadata, we used the PyMuPDF (fitz) library to extract annotations associated with internal and external links. This included each link’s source text, bounding box coordinates, and its target- either a page location within the document or a URL. For semantic tag structure, including headings, lists, and table tags, we employed the Pdfix SDK, a PDF/UA-compliant parser that provides access to the document’s complete tag tree.

To simulate a screen reader user experience, we developed a custom automation pipeline using AppleScript to activate macOS VoiceOver, open each document in Preview, and issue the “Read All” command. The audio output was recorded using QuickTime and transcribed into text using Amazon transcribe. Although these transcripts surfaced rich semantic cues such as “Heading Level 2” and “Table'', they were often noisy or incomplete. To supplement this, we also extracted the programmatic reading order by aligning Marked Content Identifiers (MCIDs) from the tag structure with rendered content.

Each data point in the final dataset, comprising 125 PDF documents, includes a reference to the modified PDF document, its accessibility label (ground truth), and associated structured inputs. The dataset is available in an open-source repository with comprehensive documentation, including input processing transcripts and pipeline setup instructions, serving as a community benchmark for future work on AI-assisted accessibility evaluation.

\begin{table}[t]
\small  
\centering
\begin{tabular}{L{28mm}L{47mm}}  
\toprule
\textbf{Criterion} & \textbf{Input Data} \\
\midrule
Alternative Text Quality & Page image, tag structure$^*$, figure(s) images, alt text(s)$^*$ \\
\addlinespace
Semantic Tagging & Page images, complete tag structure$^*$, screen reader transcript$^*$ \\
\addlinespace
Logical Reading Order & Page images, document reading order$^*$, screen reader transcript$^*$ \\
\addlinespace
Table Structure & Page image, table(s) images, table tags$^*$ structure(s) \\
\addlinespace
Functional Hyperlinks & Page images, link metadata (internal and external)$^*$, tag structure$^*$ \\
\addlinespace
Color Contrast & Page image, foreground and background color values$^*$ \\
\addlinespace
Font Readability & Page image, font metadata (font size, type, text content, embedding)$^*$ \\
\bottomrule \\ [-3mm]
\multicolumn{2}{c}{}$^*$The input is withheld for papers with the \textit{Cannot Tell (CT)} label. \\
\end{tabular}
\caption{Input data provided to model to evaluate each accessibility criterion}
\label{tab:input-data}
\end{table}


\section{Benchmarking LLMs for Accessibility Evaluation}
\label{sec:experiments}
This section details our framework for evaluating LLMs capabilities as PDF accessibility evaluators using our benchmark dataset. We outline our model selection, prompt design process, and evaluation protocol.

\subsection{Model Selection}
We evaluated five state-of-the-art models representing different capabilities (text-only versus multimodal), architectures (proprietary and open-source), training methodologies, and context window sizes: GPT-4-Turbo and GPT-4o-Vision from OpenAI, Claude-3.5 from Anthropic, Gemini-1.5-Pro from Google, and LLaMA-3.2-Vision-Instruct from Meta. GPT-4o and Gemini-1.5-Pro are fully multimodal, capable of processing both text and visual inputs, while GPT-4-Turbo and Claude-3.5 are text-focused. LLaMA-3.2-Vision-Instruct offers open-weight multimodal capabilities.

These models vary in context window size, ranging from 128K to 1 million tokens. For consistent evaluation, we used the most recent stable version of each model available as of February 2025. All models were accessed through their respective official APIs using default configuration settings (temperature=0.7) without any fine-tuning. 

\subsection{Prompt Design and Optimization}
We chose a zero-shot prompting approach as it reflects current LLM capabilities and addresses practical constraints -- there are no large, labeled training datasets available for fine-tuning models on PDF accessibility evaluation tasks. We use this setup to evaluate each model along each accessibility criterion. We developed and applied a structured, step-by-step prompt template, inspired by \citet{lopez}, across all models; the prompt included five components:
\begin{itemize}
    \item Task definition: Establishes the model's role as an accessibility evaluator and specifies the criterion being assessed
    \item Guidelines referenced: Provides relevant WCAG 2.2 and PDF/UA guidelines for the criterion
    \item Sub-criteria checklist: Provides the specific sub-criteria list corresponding to the criterion (example in Table~\ref{tab:criteria-summary})
    \item Output label definitions: Defines the four accessibility labels (Passed, Failed, Not Present, Cannot Tell)
    \item Output format instructions: Requests a structured JSON response containing: (a) the accessibility label, (b) identified issues with explanations (if any), and (c) remediation recommendations for addressing any issues.
\end{itemize}
The prompt requests for both a high-level accessibility label and assessment by each sub-criterion. For instance, when evaluating alternative text quality, a model might assign a "Failed" label while identifying specific issues such as "Alt text for Figure 3 is insufficient to understand the context" and providing actionable recommendations like "Include chart's key data trends to improve the alt text." A sample prompt and example output for alternative text quality evaluation is included in Appendix~\ref{sec:prompt-output}.

We developed our prompts through an iterative refinement process that involved multiple stages of testing. We first created baseline prompts incorporating accessibility standards and evaluation requirements based on WCAG and PDF/UA guidelines. We then tested them on a subset of documents from the benchmark dataset and analyzed response patterns to identify areas for improvement. Based on these pilot results, we refined the prompts to clarify ambiguous instructions, add explicit guidance for edge cases, modify accessibility label definitions for each criterion, and standardize output formatting requirements. 
This optimization process yielded prompts that consistently generated structured and consistent output responses across all models.

\subsection{Accessibility Evaluation}
For each document in our benchmark, we tested each model against each criterion by supplying the criterion-specific prompt along with the corresponding input data (as described in Section~\ref{sec:document_representationa}). From the models' structured JSON responses, we extracted three key elements: the primary accessibility label (Passed, Failed, Not Present, or Cannot Tell), identified accessibility issues with explanations, and recommended remediation steps for each identified issue.


Due to context window limitations, even with models capable of handling large inputs, many scholarly PDFs with several complex visual elements and tag structures exceeded these limits. As a result, we processed documents page-by-page, requiring us to consolidate multiple page-level evaluations into a single document-level accessibility label. When different pages of the same document received different labels (e.g., Page 1 "Passed" but Page 2 "Failed"), we applied the most conservative label to the entire document, considering it "Passed" only if all its pages passed. This page-by-page evaluation was particularly relevant for criteria with potentially uneven distribution throughout documents, such as alternative text quality and table structure. Though some criteria like font readability and color contrast tend to remain more consistent throughout documents. 

For consistency, we applied this conservative labeling approach across all criteria, not only at the page level but also at the sub-criteria level. The structured JSON responses sometimes included the model's assessment of each sub-criterion (e.g., "descriptiveness," "clarity," "conciseness" for alternative text quality). 
If the model identified failure in any sub-criterion (e.g., "Alt text lacks details for Figure 3"), the entire document was labeled as "Failed" for that accessibility criterion, regardless of other sub-criteria compliance. 

To account for potential variability in model outputs, each accessibility evaluation was repeated three times with identical inputs. We calculated the mean accuracy for each model-criterion combination along with the standard deviation (SD) to assess not only performance but also consistency across evaluations.

\section{Benchmarking Results and Analysis}
\label{sec:results}

\begin{figure*}[t!]
  \centering
    \includegraphics[width=1\linewidth]{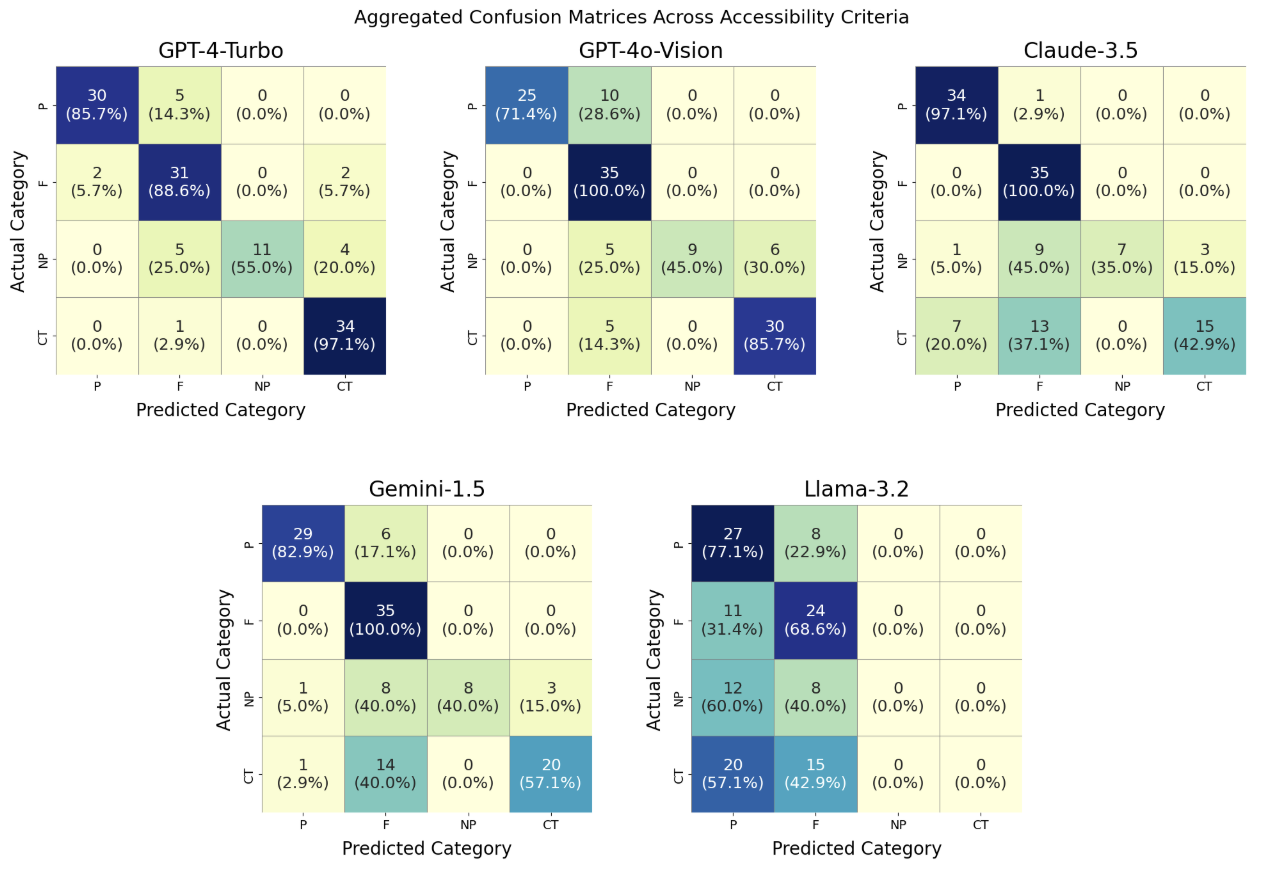}
  \caption{Aggregated Confusion Matrices across accessibility criteria}
  \label{f1_scores}
  \Description{Grid of five 4×4 confusion matrices showing the aggregate prediction performance of five language models—GPT-4 Turbo, GPT-4o Vision, Claude-3.5, Gemini-1.5, and Llama-3.2—across all criteria on accessibility label classification. Each matrix plots actual versus predicted labels: Passed (P), Failed (F), Not Present (NP), and Cannot Tell (CT).
  Each cell includes the aggregate count of predictions across all criteria and the corresponding percentage, indicating how well each model identified the true label. GPT-4 Turbo shows high diagonal values, especially for CT (34 correct, 97.1\%) and F (31, 88.6\%). GPT-4o Vision also performs well but misclassifies some Passed documents as Failed (10 instances, 28.6\%). Claude-3.5 has strong performance on Passed (97.1\%) and Failed (100\%) but struggles with Not Present and Cannot Tell, often confusing them with Failed. Gemini-1.5 performs well on P and F but shows dispersion across CT and NP labels. Llama-3.2 performs the worst overall, failing to predict any instances of NP and CT, often misclassifying their instances as Passed or Failed.
  }
\end{figure*}


This section presents the findings from our evaluation of five LLMs---GPT-4-Turbo, GPT-4o-Vision, Claude-3.5 Sonnet, Gemini-1.5-Pro, and LLaMA-3.2-Vision-Instruct---across seven PDF accessibility criteria. We examine overall performance metrics, criterion- and accessibility label-specific results, and model-specific behavior patterns. In Section~\ref{sec:comparison}, we also provide a qualitative comparative analysis with traditional automated checkers, which generally cannot assess the criteria included in our benchmark dataset.

\subsection{Overall Performance Comparison}
\begin{table*}[t]
\small
\centering
\begin{tabular}{lccccc}
\toprule
\textbf{Criterion} & \textbf{GPT-4 Turbo} & \textbf{GPT-4o Vision} & \textbf{Claude 3.5} & \textbf{Gemini 1.5} & \textbf{Llama 3.2} \\
\midrule
Alt Text Quality         & 0.70 $\pm$\ 0.04 & 0.50 $\pm$\ 0.00 & 0.50 $\pm$\ 0.00 & 0.50 $\pm$\ 0.00 & 0.40 $\pm$\ 0.04\\
Semantic Tagging         & 0.85 $\pm$\ 0.04 & 0.85 $\pm$\ 0.04 & 0.90 $\pm$\ 0.04 & 0.80 $\pm$\ 0.08 & 0.35 $\pm$\ 0.08 \\
Logical Reading Order    & 0.67 $\pm$\ 0.11 & 0.87 $\pm$\ 0.05 & 1.00 $\pm$\ 0.00 & 0.93 $\pm$\ 0.00 & 0.27 $\pm$\ 0.05 \\
Table Structure       & 1.00 $\pm$\ 0.00 & 0.75 $\pm$\ 0.00 & 0.55 $\pm$\ 0.04 & 0.55 $\pm$\ 0.04 & 0.45 $\pm$\ 0.04 \\
Functional Hyperlinks      & 0.80 $\pm$\ 0.04 & 0.75 $\pm$\ 0.04 & 0.80 $\pm$\ 0.04 & 0.85$\pm$\ 0.04 & 0.50 $\pm$\ 0.00 \\
Color Contrast   & 0.93 $\pm$\ 0.05 & 1.00 $\pm$\ 0.00 & 0.67 $\pm$\ 0.00 & 0.60 $\pm$\ 0.00 & 0.53 $\pm$\ 0.00 \\
Font Readability & 1.00 $\pm$\ 0.00 & 0.93 $\pm$\ 0.00  & 0.73 $\pm$\ 0.00 & 1.00 $\pm$\ 0.00 & 0.47 $\pm$\ 0.04\\
\midrule
\textbf{Average} & \textbf{0.85 $\pm$\ 0.04} & \textbf{0.81 $\pm$\ 0.02} & \textbf{0.74 $\pm$\ 0.02} & \textbf{0.75 $\pm$\ 0.02} & \textbf{0.42 $\pm$\ 0.04} \\
\bottomrule

\end{tabular}
\caption{Per-criterion accuracy across models (Mean ± Standard Deviation)}
\label{tab:criterion-accuracy}
\end{table*}

\begin{table*}[t]
\centering
\small
\begin{tabular}{L{28mm}ccccccccccccccc}
\toprule
\multirow{2}{*}{\textbf{Criterion}} & \multicolumn{3}{c}{\textbf{GPT-4 Turbo}} & \multicolumn{3}{c}{\textbf{GPT-4o Vision}} & \multicolumn{3}{c}{\textbf{Claude 3.5}} & \multicolumn{3}{c}{\textbf{Gemini 1.5}} & \multicolumn{3}{c}{\textbf{Llama 3.2}} \\
\cmidrule(lr){2-4} \cmidrule(lr){5-7} \cmidrule(lr){8-10} \cmidrule(lr){11-13} \cmidrule(lr){14-16}
& P & R & F1 & P & R & F1 & P & R & F1 & P & R & F1 & P & R & F1 \\
\midrule
Alt Text Quality          & 0.57 & 0.70 & 0.62 & 0.33 & 0.50 & 0.38 & 0.33 & 0.50 & 0.38 & 0.33 & 0.50 & 0.38 & 0.22 & 0.40 & 0.28 \\
Semantic Tagging          & 0.91 & 0.85 & 0.84 & 0.91 & 0.85 & 0.84 & 0.93 & 0.90 & 0.90 & 0.89 & 0.80 & 0.80 & 0.18 & 0.35 & 0.23 \\
Logical Reading Order    & 0.64 & 0.67 & 0.65 & 0.91 & 0.87 & 0.86 & 1.00 & 1.00 & 1.00 & 0.94 & 0.93 & 0.93 & 0.18 & 0.27 & 0.21 \\
Table Structure       & 1.00 & 1.00 & 1.00 & 0.63 & 0.75 & 0.67 & 0.59 & 0.55 & 0.46 & 0.76 & 0.55 & 0.54 & 0.17 & 0.35 & 0.22 \\
Functional Hyperlinks  & 0.89 & 0.80 & 0.76 & 0.85 & 0.75 & 0.71 & 0.87 & 0.80 & 0.75 & 0.89 & 0.85 & 0.83 & 0.25 & 0.50 & 0.34 \\
Color Contrast    & 0.94 & 0.93 & 0.93 & 1.00 & 1.00 & 1.00 & 0.76 & 0.67 & 0.61 & 0.43 & 0.60 & 0.49 & 0.35 & 0.53 & 0.42 \\
Font Readability  & 1.00 & 1.00 & 1.00 & 0.94 & 0.93 & 0.93 & 0.85 & 0.73 & 0.68 & 1.00 & 1.00 & 1.00 & 0.30 & 0.47 & 0.36 \\
\midrule
\textbf{Average} & \textbf{0.85} & \textbf{0.85} & \textbf{0.83} & \textbf{0.79} & \textbf{0.81} & \textbf{0.77} & \textbf{0.85} & \textbf{0.74} & \textbf{0.68} & \textbf{0.75} & \textbf{0.75} & \textbf{0.71} & \textbf{0.23} & \textbf{0.41} & \textbf{0.29} \\

\bottomrule
\end{tabular}
\caption{(Macro-) Precision (P), Recall (R) and F1 scores across models for each criterion}
\label{tab:macro-metrics}
\end{table*}

We observed significant variations in model performance across different accessibility criteria and labels. Table~\ref{tab:criterion-accuracy} presents the mean accuracy and standard deviations for each model-criterion combination computed across three evaluation runs. Among the models, GPT-4-Turbo achieved the highest average accuracy (0.85), while Llama-3.2 demonstrated the lowest (0.42). Macro-averaged precision, recall, and F1 scores are reported in Table~\ref{tab:macro-metrics}. Across all models, alt text quality is one of the most difficult criterion for evaluation, while several models struggled also with table structure and color contrast. Aggregated confusion matrices for each model are presented in Figure~\ref{f1_scores}. 

We highlight the following trends across models and criteria:
\begin{itemize}[leftmargin=16pt]
    \item Models exhibit accessibility label-specific evaluation capabilities. All models performed substantially better on Passed and Failed labels (where features are present) than on Not Present and Cannot Tell labels (where features are absent or information provided is insufficient to produce a judgment). 
    This pattern is evident across all criteria, suggesting a fundamental limitation in current LLMs' ability to recognize when evaluation is not applicable or impossible. Among the models, GPT-4-Turbo led in recognizing when evaluated features were absent (55\% accuracy for Not Present) and when information was insufficient (97.1\% accuracy for Can't Tell), while other models struggled much more with these label classes.
    \item Some models show differential performance between structure-focused criteria (semantic tagging, logical reading order, table structure) versus visually-focused criteria (color contrast, alternative text, font readability). Claude-3.5 showed better performance on structural evaluation (mean accuracy 82\%) compared to visual assessment (mean accuracy 63\%), while GPT-4o-Vision performs very well on visually-oriented criteria like color contrast (perfect accuracy). Llama-3.2 shows systematic weaknesses across all criteria.
    \item Models exhibit some distinctive error patterns, with Claude-3.5 and Gemini-1.5 showing a tendency to hallucinate content when information was missing. For instance, when not provided with extracted alt text as input, Claude frequently generated an alt text description from scratch and then evaluated it. GPT-4o-Vision demonstrated heightened sensitivity to minor issues, such as flagging scholarly citation links as accessibility failures despite expert evaluation to the contrary. GPT-4-Turbo performed particularly well in cases with information gaps (97.1\% accuracy on Cannot Tell scenarios).
\end{itemize}

\begin{table*}[t!]
\centering
\small
\begin{tabular}{lccccc}
\toprule
\textbf{Accessibility label} & \textbf{GPT-4 Turbo} & \textbf{GPT-4o Vision} & \textbf{Claude 3.5} & \textbf{Gemini 1.5} & \textbf{Llama 3.2} \\
\midrule
Passed        & 0.86 $\pm$\ 0.04 & 0.71 $\pm$\ 0.07 & 0.97 $\pm$\ 0.00 & 0.83 $\pm$\ 0.02 & 0.77 $\pm$\ 0.05\\
Failed        & 0.89 $\pm$\ 0.04 & 1.00 $\pm$\ 0.00 & 1.00 $\pm$\ 0.00 & 1.00 $\pm$\ 0.00 & 0.69 $\pm$\ 0.08 \\
Not Present   & 0.55 $\pm$\ 0.04 & 0.45 $\pm$\ 0.00 & 0.35 $\pm$\ 0.04 & 0.40 $\pm$\ 0.04 & 0.00 $\pm$\ 0.00 \\
Cannot Tell   & 0.97 $\pm$\ 0.02 & 0.86 $\pm$\ 0.00  & 0.43 $\pm$\ 0.02  & 0.57 $\pm$\ 0.02 & 0.00 $\pm$\ 0.00\\
\bottomrule
\end{tabular}
\caption{Label-Specific Accuracy by Model}
\label{tab:category_accuracy}
\end{table*}


\subsection{Criterion-Specific Performance Analysis}

We now summarize and discuss model performance for each accessibility criterion. 

\subsubsection{Alternative Text Quality}
This criterion presented challenges for all models, with GPT-4-Turbo achieving the highest accuracy (0.70) while other models performed considerably lower (0.40-0.50). No model except GPT-4o correctly identified a single document with the `Not Present' and `Cannot Tell' accessibility labels.


A revealing pattern appeared across all models except GPT-4-Turbo: they consistently misclassified documents without images (`Not Present' label) as `Failed', citing "missing alternative text" as the violation. This misclassification indicates that models interpreted the absence of alt text as a failure even when there are no images, and even though our objective was to check if LLMs can evaluate the quality of alt text rather than their presence. 
Claude-3.5 occasionally hallucinated alt text descriptions for images without alt text and then marked the document as Failed based on those hallucinated descriptions. For the `Cannot Tell' label, all models except GPT-4-Turbo incorrectly classified documents as `Failed', also due to predictions based on hallucinated descriptions generated by those models.

\subsubsection{Semantic Tagging}
Claude-3.5 demonstrated the highest accuracy (0.90) followed by GPT-4-Turbo and GPT-4o-Vision (both 0.85). We found that models occasionally misclassified `Passed' documents as `Failed' due to detecting minor structural issues that we did not consider significant enough to warrant failure, i.e., models applied more stringent interpretations of guidelines than human annotators. For instance, GPT-4o-Vision flagged documents with issues such as "Replace non-semantic <Span> tags with appropriate structural tags", even when the <Span> tags had been appropriately marked as per guidelines. 

Llama-3.2 was the only model that could not correctly identify any Not Present cases. 
For the `Cannot Tell' label, GPT-4 models achieved perfect accuracy on our evaluation set, while Claude-3.5 and Gemini 1.5 labeled some documents as Failed, due to identifying some missing tags. Llama-3.2 also struggled significantly with this criterion and also did not provide any explanation or recommendations as instructed in the prompt.


\subsubsection{Logical Reading Order}
All models achieved high performance for this criteria, with Claude-3.5 achieving perfect accuracy (1.00) and Gemini-1.5 nearly matching it (0.93). For the `Cannot Tell' label, all models except Llama-3.2 correctly identified that there was insufficient information. GPT-4-Turbo and Llama-3.2 models occasionally misclassified documents with `Passed' and `Failed' labels in both directions, especially when documents contained artifacts that were assigned an incorrect reading order.

\subsubsection{Table Structure}
GPT-4-Turbo achieved perfect accuracy (1.00) on this criterion, while other models showed moderate performance (0.45-0.75). 
For the `Not Present' label, Claude-3.5, Gemini-1.5, and Llama-3.2 models frequently misclassified documents without tables as `Failed', misinterpreting some elements like figures containing sub-figures, comment boxes, and notes as tables with missing tags. For the `Cannot Tell' label, the GPT-4 models correctly identified all instances, while other models frequently misclassified them as `Failed'. 

\subsubsection{Functional Hyperlinks}
This criterion showed high accuracy across all models (0.75-0.85) except Llama (0.50). GPT-4o-Vision demonstrated heightened sensitivity to descriptiveness in internal links and raised concerns such as "Some links lack descriptive text", frequently flagging `Passed' documents as `Failed' even when links met basic accessibility requirements as per manual evaluation (considering scholarly documents usually don't have descriptive links). This suggests that different models apply varying thresholds when interpreting accessibility guidelines for scholarly content, where citation links typically follow domain-specific conventions. For the `Cannot Tell' label, most models correctly recognized it with 100\% accuracy.

\subsubsection{Color Contrast}
GPT-4o-Vision achieved perfect accuracy (1.00) on this criterion, followed by GPT-4-Turbo (0.93). This likely stems from GPT-4o's enhanced visual processing capabilities, which are particularly valuable for evaluating contrast ratios. The main challenge for lower-performing models, namely Claude-3.5, Gemini-1.5, and Llama-3.2, appeared in the `Cannot Tell' label, where Claude-3.5 and Gemini-1.5 frequently misclassified these documents based on hallucinated color codes rather than recognizing information insufficiency. Interestingly, we found that Claude-3.5 tended to default to `Passed' in uncertain cases, while Gemini-1.5 more often defaulted to `Failed', revealing different underlying biases in models when facing uncertainty.

\subsubsection{Font Readability}
This criterion showed strong performance from GPT-4-Turbo and Gemini-1.5 (both having accuracy 1.0), followed closely by GPT-4o-Vision (0.93). The primary challenge was again in identifying the `Cannot Tell' label, where Claude-3.5 frequently misclassified documents as `Passed' even when font metadata (e.g., size, type, and embedding) was deliberately withheld, based on visual perception and hallucinated metadata. This tendency to default to `Passed' when lacking information suggests an underlying bias toward assuming compliance, which is concerning for an accessibility checker tool. However, Llama-3.2's performance (0.47) showed significant confusion, particularly in distinguishing between `Failed and `Cannot Tell' labels, often defaulting to `Failed' without providing any explanation.

Our analysis reveals concerning hallucination patterns that varied by accessibility criterion, with significant implications for evaluation reliability. Most notably, Claude-3.5 fabricated alt text descriptions in 32\% of "Cannot Tell" cases, generating nonexistent descriptions and then evaluating them for accessibility. Similarly, for font readability assessment, Claude-3.5 frequently hallucinated font metadata when this information was deliberately withheld, defaulting to "Passed" classifications based on visual appearance rather than recognizing information insufficiency. GPT-4o-Vision occasionally generated color values for contrast evaluation, while Gemini-1.5 showed similar patterns across multiple criteria. These patterns undermine the reliability of LLM-based accessibility evaluation, particularly for compliance workflows requiring reproducible results.



\subsection{Model-Specific Evaluation Patterns}


As shown in Table~\ref{tab:category_accuracy}, models demonstrate varying performance across accessibility labels. 
Some models exhibit low variance in performance across labels (e.g., Claude-3.5), while others exhibit much higher variance (e.g., GPT-4-Turbo, Llama-3.2). Differential performance across labels could impact how models perform in real-world settings with biased label distributions. 

We also identified distinct error biases when models face uncertainty (when insufficient information is provided). Claude-3.5 exhibited a tendency to default to `Passed' in uncertain cases, particularly evident in font readability assessment where it frequently misclassified `Cannot Tell' scenarios as compliant. In contrast, Gemini-1.5 showed a bias toward labeling uncertain cases as `Failed.' These opposing biases have meaningful implications for how these models should be deployed: Claude's optimistic bias might reduce false positives but increase false negatives, while Gemini's more conservative bias presents the opposite trade-off.

Our results show that models excel at specific accessibility criteria. Claude-3.5 performed best on structure-focused evaluation, achieving perfect accuracy (1.00) on logical reading order and strong performance on semantic tagging (0.90), suggesting particular strength in understanding document hierarchies and content relationships. In contrast, GPT-4o-Vision demonstrated its capabilities in visually-oriented criteria, achieving perfect accuracy (1.00) on color contrast evaluation and strong performance (0.93) on font readability assessment, likely leveraging its enhanced visual processing capabilities. Gemini-1.5 showed balanced capabilities across both structural and visual criteria, with particularly strong performance on reading order (0.93) and font readability (1.00).

GPT-4-Turbo's strong performance in recognizing information insufficiency (97.1\% accuracy on `Cannot Tell') positions it as particularly valuable for preliminary screening, where acknowledging evaluation limitations is as important as detecting any issues. These distinct behavioral patterns suggest that model selection should be criterion-specific rather than using one model for all assessments. For comprehensive accessibility evaluation, optimal results might be achieved through strategically combining models based on their demonstrated strengths and reliability.

\subsection{Analyzing Remediation Recommendations}


We conducted a systematic qualitative analysis of remediation recommendations provided by models for documents classified as 'Failed', covering a sample of 45 recommendations (9 from each model) across different accessibility criteria. Our analysis was conducted by one of the authors with expertise in accessibility evaluation, and each recommendation was evaluated using the following rubric:
\begin{itemize}[leftmargin=16pt]

\item Specificity: Whether issues were location-aware and referenced specific document elements (e.g., "Table 1, page 3" rather than generic "tables should be properly formatted")

\item Actionability: Whether concrete, implementable remediation steps were provided (e.g., "Add scope='col' to header cells" rather than "improve table structure")

\item Technical accuracy: Whether recommendations aligned with established WCAG/PDF-UA standards and represented valid solutions to identified issues
\end{itemize}


GPT-4-Turbo and GPT-4o-Vision consistently (in > 85\% examined outputs) provided the most specific and technically accurate recommendations, including references to specific document elements requiring remediation. For example, in table structure evaluation, GPT-4-Turbo identified specific tables and cells needing header associations rather than giving general advice:
\begin{quote}
\small
\texttt{Table 1 on page 3 lacks proper header associations. Add TH tags to the first row cells and set scope='col' attribute. For the leftmost column, set scope='row' for the header cells.}
\end{quote}

\noindent On the other hand, Claude-3.5 and Gemini-1.5 tended to provide technically sound but less specific recommendations (in approximately 60-70\% outputs) that would require additional investigation by document authors, e.g.,:
\begin{quote}
\small
\texttt{Ensure all tables have proper header cells. Use TH tags for header cells and ensure they have appropriate scope attributes.}
\end{quote}

\noindent Llama-3.2 did not provide relevant recommendation and actionable steps. Sample outputs showing high-quality and low-quality recommendations for table structure criterion by GPT-4-Turbo and Gemini-1.5 models respectively, are presented in Appendix~\ref{sec:recommendation-outputs}. 

This analysis highlights that remediation guidance quality represents an important dimension beyond simple classification accuracy for practical accessibility evaluation tools. While a comprehensive analysis of remediation recommendations extends beyond this paper's scope, our findings suggest significant variability in how effectively different models translate accessibility evaluations into actionable guidance. 


\section{Qualitative Comparison with Automated Evaluation Tools}
\label{sec:comparison}
To contextualize our LLM-based evaluation findings, we conducted a qualitative comparison with five leading automated PDF accessibility checkers: Adobe Acrobat Pro, PAC 2024, PAVE, axesPDF, and CommonLook PDF Validator. We analyzed a subset of our benchmark dataset (35 documents across seven criteria) using all five checkers, and identified
differences in evaluation capabilities, strengths, and limitations that inform how these different approaches might complement each other as well as LLM-based evaluation.

\subsection{Automated Checkers Overview}

\begin{table*}[ht]
\centering
\resizebox{\textwidth}{!}{%
\begin{tabular}{|>{\raggedright\arraybackslash}p{3.2cm}|>{\raggedright\arraybackslash}p{2.8cm}|>{\raggedright\arraybackslash}p{2.8cm}|>{\raggedright\arraybackslash}p{2.5cm}|>{\raggedright\arraybackslash}p{2.5cm}|>{\raggedright\arraybackslash}p{2.8cm}|>{\raggedright\arraybackslash}p{3.6cm}|}
\hline
\rule{0pt}{2.5ex} 
\textbf{Criterion} & \textbf{Adobe Acrobat Pro} & \textbf{PAC 2024} & \textbf{PAVE} & \textbf{axesPDF} & \textbf{CommonLook} & \textbf{LLM-based Approaches} \\
\hline
\rule{0pt}{2.5ex} \textbf{Alternative Text} & Verifies presence but cannot assess quality or appropriateness. Doesn't identify decorative images properly. & Verifies presence and checks for common issues (e.g., file paths as alt text). Warns if non-figure elements have alt text. & Basic presence check only. Cannot evaluate quality or relevance. & Similar to PAC 2024. Can detect figures even when not properly tagged. & Comprehensive presence check but also provides tools for manual assessment of appropriateness. Includes validation against multiple standards. & Can evaluate semantic quality and contextual appropriateness of alt text. May hallucinate content for Cannot Tell scenarios. \\
\hline
\rule{0pt}{2.5ex} 
\textbf{Semantic Tagging} & Confirms document is tagged but doesn't verify tag correctness or semantic appropriateness. AutoTag feature attempts to determine correct tags. & Evaluates formal correctness and warns about inappropriate structure elements. Provides tag tree view with properties and attributes. & Checks document tagging and some nesting rules. Limited semantic verification though highlights untagged elements. & Similar to PAC 2024. Warns if document lacks heading tags and evaluates tag tree structure. & Detailed structural validation including proper nesting. Provides more comprehensive semantic assessment tools through specific wizards. & Effectively evaluates semantic appropriateness of tags and can identify logical structural issues. Sometimes applies overly strict interpretations. \\
\hline
\rule{0pt}{2.5ex} 
\textbf{Logical Reading Order} & Always marked as "Needs Manual Check." Only verifies tab order consistency with structure. & Checks tab order and basic tagging sequence. No visual vs. programmatic order validation. & Visual inspection via tab navigation only. No automated assessment. & Similar to PAC 2024. Provides screen reader preview to help assess reading order. & Verifies technical reading order structure but requires user verification for logical assessment. Provides tools to set proper tabbing order to follow document structure. & Strong performance in assessing logical reading order coherence. Some models (Claude 3.5) achieve perfect accuracy matching expert judgment. \\
\hline
\rule{0pt}{2.5ex} 
\textbf{Table Structure} & Examines basic structure (headers, regularity) but cannot assess header-data cell associations. & Checks for proper table tags and warns if tables lack headers. Limited assessment of semantic relationships. & Very basic header presence check. No validation of structure or associations. & Similar to PAC 2024. Cannot detect incorrect header-data relationships. & Comprehensive technical validation of table markup. Limited semantic assessment. & Can assess whether header-data relationships make semantic sense. Identifies when column headers semantically mismatch their data. \\
\hline
\rule{0pt}{2.5ex} 
\textbf{Hyperlinks} & Checks for repetitive links. Doesn't evaluate activity, understandability, or descriptiveness. & Verifies link tag presence. Expects alt text for links. No evaluation of link purpose or descriptiveness. & Basic checks for link properties. Limited assessment of link descriptiveness. & Similar to PAC 2024. Can check if linkable text contains proper link tags with necessary attributes. & Verifies hyperlinks are not broken. Limited assessment of descriptiveness. & Evaluates link descriptiveness and contextual clarity. Sometimes overly strict on scholarly link text conventions. \\
\hline
\rule{0pt}{2.5ex} 
\textbf{Color Contrast} & Always requires manual check. No automated assessment. & Can identify poor contrast but often marked as "Not Applicable." Relies on user verification. & No automated check for contrast. Requires manual verification. & Similar to PAC 2024. Can flag "invisible" text (e.g., white-on-white) but limited contrast analysis. & Includes contrast checks but requires user verification for final assessment. & Strong performance in contrast evaluation, especially multimodal models (GPT-4o achieved perfect accuracy). \\
\hline
\rule{0pt}{2.5ex} \textbf{Font Readability} & Checks character encoding only. No assessment of size, style, or spacing. & More strict embedding checks. Classifies font types but no readability assessment. & No automated checks for font properties beyond basic encoding verification. & Similar to PAC 2024. More stringent on embedding requirements. & Verifies embedded fonts but limited evaluation of readability impacts. & Can assess both technical compliance and readability impact of font choices. Multiple models achieved perfect accuracy. \\
\hline

\end{tabular}
}
\caption{Comparison of Accessibility Evaluation Tools and LLM-Based Approaches}
\label{tab:accessibility_comparison}
\end{table*}

Table~\ref{tab:accessibility_comparison} summarizes our qualitative findings across the seven accessibility criteria. Our comparison revealed that automated checker tools operate primarily as technical compliance verifiers rather than accessibility evaluators, i.e., they identify structural issues effectively but struggle with contextual and semantic assessment. 

Adobe Acrobat Pro, despite being the industry standard, employs a relatively simple three-category classification system (Passed / Failed / Needs Manual Check) across 32 criteria. For criteria requiring semantic judgment, such as reading order and color contrast, Acrobat Pro consistently defers to "Needs Manual Check," which minimizes false positives but results in many indeterminate evaluations requiring human intervention.

PAC 2024 and axesPDF implement more granular evaluations with four accessibility labels (Passed / Failed / Warned / Not Applicable) across 40 criteria spanning PDF/UA, WCAG, and quality checks. These tools demonstrate greater specificity but also exhibit a higher tendency toward false negatives. For instance, when evaluating our documents that `Passed' the semantic tagging criterion, PAC 2024 frequently failed documents when decorative elements like table borders were not tagged, despite accessibility guidelines explicitly exempting such elements. This example illustrates how automated checkers sometimes prioritize formal compliance over functional accessibility.

PAVE, with its focus on EU Standards, offers the most limited evaluation scope among the tested tools, with no comprehensive criteria list and primarily emphasizing document metadata and structural tagging. Its minimal checking approach makes it quick to use but inadequate for comprehensive accessibility evaluation. At the other extreme, CommonLook PDF Validator provides the most extensive coverage with 313 distinct accessibility checks, but its complexity introduces significant usability barriers, requiring a large amount of time to process a single document and considerable expertise to interpret results effectively.


\subsection{Comparative Strengths and Limitations: LLMs and Automated Checkers}

Automated checkers demonstrate significant advantages in technical verification, applying consistent rule-based assessment at scale. These tools process documents rapidly, typically completing evaluations in seconds, and provide deterministic results for well-defined structural requirements. They excel in detecting missing required elements, such as document tags, language specifications, or alt text fields. When evaluating font embedding requirements, both PAC 2024 and axesPDF correctly identified all instances of non-embedded fonts in our documents, providing specific information about which fonts required embedding. Such technical evaluation makes these tools particularly valuable for initial compliance screening and identifying fundamental structural barriers.

However, automated checkers frequently produce false negatives by flagging technical violations that did not constitute meaningful accessibility barriers. This was particularly evident with PAC 2024 and axesPDF, which routinely failed documents when spaces between words were not tagged explicitly, despite these elements having no functional impact on screen reader navigation or content comprehension. Another notable limitation was their inability to handle `Not Present' scenarios appropriately. For documents containing no tables or figures, most checkers still provided pass or fail judgments for these criteria, applying binary assessments where a more nuanced approach was needed. This approach fails to distinguish between criteria that genuinely do not apply to a document, a critical requirement in accessibility assessment.

LLM-based approaches, in contrast, could be used to conduct semantic evaluation, a capability that automated checkers lack. This strength was evident in our reading order evaluation, where Claude-3.5 achieved perfect accuracy (100\%) in determining whether the programmatic reading order logically followed the visual layout. This contrasts sharply with all five automated checkers, which either marked reading order evaluation as "Needs Manual Check" (Adobe Acrobat Pro) or could only verify sequence properties without assessing logical coherence (PAC 2024, axesPDF, CommonLook). LLM semantic understanding extends beyond single-element assessment to evaluating contextual relationships between document components. In table structure evaluation, while automated tools could verify the presence of header cells, they universally failed to identify semantically incorrect header-data relationships.

LLMs also provide nuanced categorization while explaining the implications of accessibility issues for readers. When evaluating semantic tagging, GPT-4-Turbo not only identified wrongly nested heading levels (H1 after H3) but explained why this structure would confuse screen reader users in the context of the document and provided specific remediation steps prioritized by severity. This level of contextual understanding and actionable guidance is in contrast to automated checkers, which typically provide generic descriptions based on rule violation rather than accessibility impact.

That said, LLMs exhibit their own limitations. Across multiple criteria, they demonstrate a tendency toward overly strict or lenient guideline interpretation, particularly for content with established conventions. This was most evident in our evaluation of navigation links, where GPT-4o-Vision frequently flagged standard academic citation links as insufficiently descriptive despite experts considering them acceptable within scholarly contexts.

LLMs also exhibit occasional hallucination behaviors that impact evaluation reliability. In alt text evaluation, for example, Claude-3.5 sometimes hallucinated alt text for images and then evaluated the accessibility of those texts rather than limiting assessment to the actual document content. Perhaps most significantly, models showed inconsistent ability to recognize when information was insufficient for evaluation, with only GPT-4-Turbo demonstrating strong performance on these `Cannot Tell' scenarios (97.1\% accuracy).

Our comparison reveals complementary strengths, demonstrating how different evaluation methods can work together to address comprehensive PDF accessibility assessment needs. For font readability and color contrast, LLM-based approaches add value in assessing visual rendering and readability assessment. LLM-based approaches can perform alt text quality evaluation, while automated checkers detect its presence, and human verification still remains valuable for complex scientific figures or charts. Table structure assessment works best with automated checkers verifying technical markup while model-based approaches can be used to assess whether header-data relationships make semantic sense. Reading order evaluation is particularly well-suited to LLM-based approaches, reducing the need for human verification except for complex layouts. For semantic tagging, automated checkers can verify tag presence, while LLM-based approaches can evaluate semantic appropriateness, with human experts resolving cases of over-stringent automated flagging. Finally, automated checkers can be used to check for structural properties like tagging of navigational links, with LLMs conducting contextual evaluation and checking for functional or broken links.


\section{Discussion}
\label{sec:discussion}

Our benchmark evaluation provides insights into the capabilities and limitations of different approaches to PDF accessibility assessment. This section discusses the contributions of our benchmark dataset and broader implications of our findings for PDF accessibility evaluation practices. Based on our experimental results, we propose a hybrid evaluation framework that leverages the complementary strengths of automated tools, LLMs, and human expertise.

\subsection{Dataset Contributions}
Our benchmark dataset addresses a significant gap in accessibility research and evaluation. Despite the critical importance of PDF accessibility for academic and professional content, prior to this work, no standardized, expert-validated dataset existed for evaluating PDF accessibility assessment approaches, which has impacted accessibility research in several key ways:

\begin{itemize}[noitemsep, topsep=1pt, leftmargin=12pt]
    \item Without standardized benchmarks, the accessibility community lacked a common reference point for comparing the effectiveness of different evaluation tools and methodologies. Automated checkers like Adobe Acrobat Pro, PAC, and axesPDF have operated with limited transparency regarding their detection capabilities across different accessibility criteria. Our benchmark provides the first comprehensive framework for measuring and comparing accessibility evaluation approaches on a common set of documents with expert-validated annotations.
    \item Our benchmark's four-category evaluation framework (Passed, Failed, Not Present, Cannot Tell) offers a more nuanced assessment than traditional binary (pass/fail) approaches. This granularity better reflects the complexity of real-world accessibility evaluation, where determining whether a criterion applies or whether sufficient information exists can be as important as detecting compliance or violations. Inclusion of accessibility labels like `Not Present' and `Cannot Tell' enables evaluation of the checker's metacognitive awareness---a critical capability for reliable assessment tools that has been largely overlooked in previous research.
    \item Our benchmark goes beyond criteria that are more easily captured by automated accessibility checkers. By incorporating seven diverse accessibility criteria aligned with WCAG and PDF/UA standards, our benchmark enables comprehensive evaluation across the spectrum of PDF accessibility guidelines. Previous studies have typically focused on isolated aspects of accessibility, such as heading structure \cite{duarte}, without providing an integrated assessment framework. Our approach reveals how evaluation methods may perform inconsistently across different accessibility criteria, with implications for developing more balanced assessment workflows in the future.
\end{itemize}

We envision our benchmark to support such research by helping developers of accessibility checker tools to identify specific strengths and weaknesses in their tools and enabling targeted improvements. Additionally, researchers can systematically compare new evaluation approaches against established baselines using standardized metrics. Finally, the benchmark can inform the development of more precise and consistent evaluation guidelines for PDF accessibility.

\subsection{Implications for PDF Accessibility Evaluation}

\begin{figure}[t!]
  \centering
    \includegraphics[width=1\linewidth]{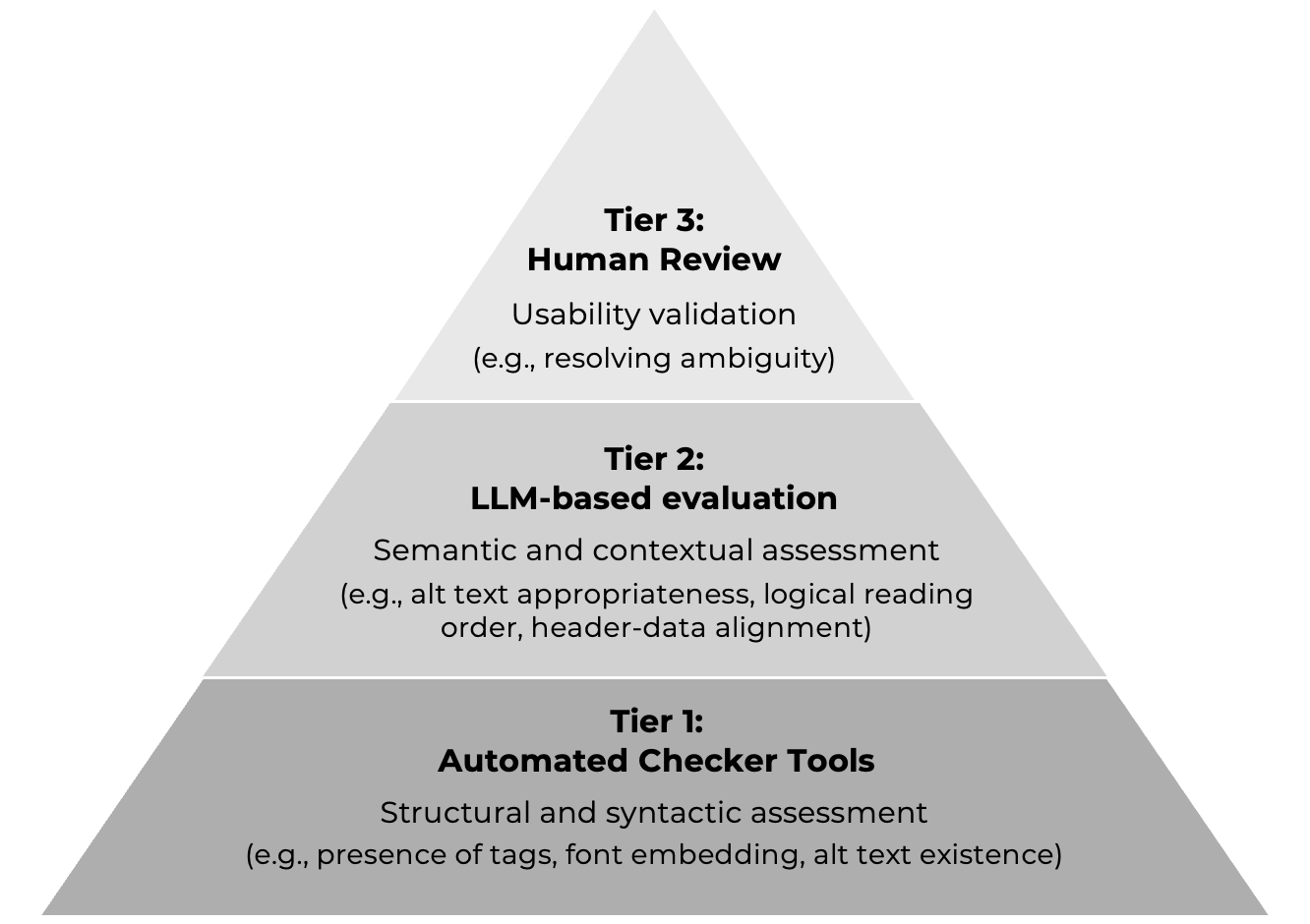}
  \caption{Three-tiered hybrid evaluation framework}
  \label{tiers}
  \Description{Three-tiered pyramid diagram illustrating the proposed hybrid workflow for PDF accessibility evaluation. The bottom tier (Tier 1: Automated Checker Tools) represents structural and syntactic assessment including presence of tags, font embedding, and alt text existence. The middle tier (Tier 2: LLM-based evaluation) covers semantic and contextual assessment such as alt text appropriateness, logical reading order, and header-data alignment. The top tier (Tier 3: Human Review) focuses on usability validation including resolving ambiguity.
  }
\end{figure}

The fundamental tension in PDF accessibility assessment involves balancing technical compliance with actual usability, and neither automated checkers nor LLM-based approaches alone fully address both these requirements. Our analysis in Section~\ref{sec:comparison} suggests that accessibility evaluation workflows should evolve beyond the current binary automated/manual paradigm toward a more nuanced, criteria-specific approach leveraging the complementary strengths of different evaluation methods. As a result, we propose a three-tiered hybrid workflow wherein by strategically combining traditional checkers, content-based evaluation, and human expertise, we can address the full spectrum of accessibility criteria more effectively than any single approach. Figure~\ref{tiers} illustrates this tiered approach, showing how each level builds upon the previous one to provide a comprehensive accessibility evaluation.

\begin{itemize}[itemsep=2pt, topsep=3pt, leftmargin=32pt]
    \item[\textbf{Tier 1:}] Traditional automated checkers would perform comprehensive technical verification, rapidly identifying fundamental issues such as missing document tags, absence of alternative text, missing language specifications, and invalid table headers. This initial screening would identify documents with major structural issues that would need to be addressed before any further evaluation.
    \item[\textbf{Tier 2:}] Employ semantic-based approaches such as LLM evaluation (similar to those introduced in this work) to assess documents that pass basic technical verification. LLM evaluation would focus on criteria requiring language understanding and contextual judgment, such as evaluating the quality and appropriateness of alternative text relative to image content, assessing logical correspondence between visual layout and reading order, verifying semantic correctness of heading structures, examining descriptiveness of link text, and evaluating meaningful relationships between table headers and data. Our analysis found that different LLMs excel at judging different criteria; for instance, Claude-3.5 showed good performance on structural assessment, while GPT-4o-Vision excelled at visual assessment. Our benchmarking methods can facilitate future evaluation of Tier 2 checkers.

    \item[\textbf{Tier 3:}] Human experts would then focus their limited time on reviewing cases where Tier 1 and 2 assessments conflict, or where Tier 2 assessments cannot determine criteria appropriateness, e.g., documents flagged with `Cannot Tell' determinations. These experts would verify critical accessibility features in high-stakes documents and conduct testing with assistive technologies. 
\end{itemize}

\noindent

This tiered approach can significantly reduce the manual evaluation burden while ensuring comprehensive assessment. By integrating these approaches, we can develop more effective and efficient accessibility evaluation systems that combine the speed and consistency of automated checking with the contextual understanding of language models and the discernment of human expertise. 


\section{Limitations and Future Work}
\label{sec:limitations}

Our dataset's scope is constrained in size and document type. With 125 PDFs focused primarily on scholarly articles, the benchmark does not fully represent the diversity of PDF documents encountered in practice, such as forms, brochures, textbooks, or reports with complex layouts. This limitation was necessary given the intensive manual effort required to create expert-validated, controlled variations across accessibility criteria, but future work should expand the benchmark to include a broader range of document types, disciplines, and sources to enhance the generalizability of our findings. Community contribution frameworks, along with semi-automated methods, could also help scale this effort of generating benchmark documents. We also advocate for the development of standardized APIs that provide programmatic access to screen reader outputs, as current methods often rely on complex workarounds.

Our evaluation focused on zero-shot prompting for language-based approaches. While this approach provides a baseline understanding of different models' capabilities without task-specific training, it likely underestimates their potential performance. Future work should explore few-shot learning, fine-tuning, and specialized model development for accessibility evaluation that could significantly improve performance and allow researchers to quantify potential gains against our zero-shot baseline.

While we chose the four-category framework to align with established WCAG-EM compliance workflows, future work can be extended to support more nuanced accessibility evaluation as the field advances. Finer-grained labels could offer additional insights into the severity and nature of accessibility issues. 


Our evaluation also did not include direct testing with users of assistive technology. Although expert validation provides a strong proxy for identifying accessibility barriers, incorporating feedback from users with disabilities would strengthen our benchmark and evaluation framework. Future work should include user studies to assess whether LLM-identified issues align with actual barriers encountered by screen reader users and other assistive technology users.

Additionally, to implement our proposed hybrid workflow in practice, future work could employ a model orchestration framework, such as LangChain, to automatically route documents through the appropriate evaluation layers. For example, an orchestration agent could first run a rule-based checker (Tier 1), then feed the output into an LLM for semantic analysis (Tier 2) if basic checks pass, and finally flag uncertain cases for human expert review (Tier 3). Such a pipeline would leverage the strengths of each component efficiently.

Finally, our work focuses only on classification accuracy rather than the quality of remediation recommendations. While we briefly analyzed the quality of remediation recommendations generated by LLMs, a thorough analysis was beyond the scope of this work. A more systematic evaluation of LLMs' capabilities to support remediation could provide valuable insights for creating more effective accessibility remediation tools.

Despite these limitations, our benchmark dataset and evaluation methodology establish a foundation for advancing PDF accessibility evaluation. Future research 
could develop hybrid evaluation approaches that combine the complementary strengths of rule-based systems and LLMs, and ultimately create more effective tools to improve document accessibility at scale.

\section{Conclusion}
\label{sec:conclusion}

In this paper, we have introduced the first expert-validated benchmark dataset for PDF accessibility evaluation spanning seven criteria derived from WCAG 2.2 and PDF/UA standards. 
Through our evaluation of five LLMs using our benchmarking framework, 
we find that different models exhibit complementary strengths, with GPT-4-Turbo achieving the highest overall accuracy (0.85), especially excelling at recognizing information insufficiency (97.1\% accuracy on `Cannot Tell'). 
However, all models struggled with the `Not Present' label and most with the `Cannot Tell' label, particularly for alt text quality assessment, indicating that models alone cannot yet replace manual assessments.

Our qualitative comparison with traditional automated checkers reveals complementary strengths: while rule-based tools excel at technical verification, LLMs better evaluate semantic appropriateness and contextual relevance. For instance, where automated checkers merely verify that alt text exists, LLMs can identify when descriptions fail to convey the image's meaning. Similarly, LLMs can assess if reading order logically follows content flow, a capability beyond current automated tools. Based on these complementary strengths, we suggest a hybrid approach combining rule-based verification with LLM-based semantic reasoning to improve PDF accessibility evaluation. Our proposed three-tiered workflow would leverage automated checkers for technical screening, LLMs for contextual judgment, and human experts for resolving edge cases, potentially reducing the manual evaluation burden while maintaining comprehensive assessment.

By establishing this benchmark dataset and evaluation framework, our work contributes to the broader goal of making scholarly content more inclusive and accessible to all readers. Future research should expand the benchmark to include a broader range of document types, explore prompting and finetuning approaches to improve LLM performance, and incorporate direct feedback from assistive technology users to further validate and refine these approaches.

\section*{Acknowledgements}

This work is partially supported by gift funds from Google and the Allen Institute for AI (Ai2), and the University of Washington eScience Institute's Azure cloud credits for research program.


\bibliographystyle{ACM-Reference-Format}
\bibliography{A11y}

\appendix
\section{Accessibility Criteria and Sub-criteria}
\label{sec:criteria_list}
This section presents the sub-criteria checklists used in prompts for evaluating corresponding accessibility criteria. These are available in the following Tables~\ref{tab:alt_text_criteria}-\ref{tab:font_criteria} corresponding to each of the seven criteria we tested.











\begin{table*}[t]
\small
\renewcommand{\arraystretch}{1.3}
\begin{tabular}{L{20mm}L{30mm}L{24mm}L{64mm}}
    \toprule
    \textbf{Criterion} & \textbf{Standards} & \textbf{Sub-criterion} & \textbf{Definition} \\
    \midrule
    \multirow[t]{6}{=}{\textbf{Alt text \newline quality}} 
    & \multirow[t]{6}{=}{%
        WCAG 2.2 SC 1.1.1 Non-text Content \newline ~\newline
        PDF/UA (ISO 14289-1) \newline ~\newline
        Best Practices}
    & Descriptiveness 
    & Alternative text fully conveys the image's purpose and content within the document context. \\
    & & Complexity Handling 
    & For complex images (charts, graphs, diagrams), alt text provides equivalent understanding of data relationships and key insights. \\
    & & Decorative Treatment
    & Images that are purely decorative are either tagged as artifacts or have empty alt text (``"). \\
    & & Clarity 
    & Text is free of unexplained jargon and redundant visual descriptions like "image of," "picture shows". \\
    & & Conciseness 
    & Alt text is concise and within recommended length limits (typically 125 characters or fewer when possible). \\
    & & Contextual Appropriateness 
    & Alt text considers surrounding content and avoids redundancy by not repeating information already presented in adjacent text. \\
    \bottomrule
\end{tabular} 
\caption{Alternative text quality and sub-criteria with associated definitions}
\label{tab:alt_text_criteria}
\end{table*}


\begin{table*}[t]
\small
\centering
\renewcommand{\arraystretch}{1.3}
\begin{tabular}{L{20mm}L{30mm}L{24mm}L{64mm}}
    \toprule
    \textbf{Criterion} & \textbf{Standards} & \textbf{Sub-criterion} & \textbf{Definition} \\
    \midrule

    \multirow[t]{5}{=}{\textbf{Semantic tagging}} 
    & \multirow[t]{5}{=}{%
        WCAG 2.2 SC 1.3.1 Info and Relationships \newline  ~\newline
        WCAG 2.2 SC 1.1.1 Non-text Content \newline  ~\newline
        WCAG 2.2 SC 2.4.2 Page Titled \newline  ~\newline
        PDF/UA (ISO 14289-1) \newline  ~\newline
        Best Practices 
    }
    & Structural Hierarchy 
    & All content elements are contained in a valid structural hierarchy. \newline Document root element \texttt{<Document>} is present and complete. \\

    & & Artifacts (Decorative Content)
    & Decorative/visual-only elements are marked as artifacts. \newline Artifact tags used properly (for page numbers, headers/footers). \\

    & & Headings
    & Heading tags (\texttt{<H1>}--\texttt{<H6>}) are used to reflect the document’s content hierarchy. \newline  Use of \texttt{<H1>} reflects the document's top-level heading structure\newline Heading levels are nested properly.\\

    & & Lists
    & List containers are marked with \texttt{<L>} tags. \newline List items are tagged with \texttt{<LI>} and include appropriate \texttt{<Lbl>} and \texttt{<LBody>} sub-elements. \newline Nested lists are structured with proper hierarchical \texttt{<L>} $\rightarrow$ \texttt{<LI>} nesting.\\

    & & Semantics (Proper Tag Usage)
    & Semantic tags are used appropriately (for instance, paragraphs are \texttt{<P>} not \texttt{<Span>}). \newline There are no empty, redundant, or irrelevant tags present in the PDF. \newline Inline tags like \texttt{<Span>}, \texttt{<Quote>}, or \texttt{<Em>} are used \textit{semantically} rather than just for styling. \newline  All structure types in use are standard or mapped to standard.  \\

    \bottomrule
\end{tabular}
\caption{Semantic tagging and sub-criteria with associated definitions}
\label{tab:tagging_criteria}
\end{table*}


\begin{table*}[!p]
\small
\renewcommand{\arraystretch}{1.3} 
\begin{tabular}{L{20mm}L{30mm}L{24mm}L{64mm}}
    \toprule
    \textbf{Criterion} & \textbf{Standards} & \textbf{Sub-criterion} & \textbf{Definition} \\
    \midrule
    \multirow[t]{8}{=}{\textbf{Logical \newline reading order}} 
    & \multirow[t]{8}{=}{%
        WCAG 2.2 SC 1.3.2 Meaningful Sequence \newline ~\newline
        WCAG 2.2 SC 2.4.3 Focus Order \newline ~\newline
        WCAG 2.2 SC 1.3.1 Info and Relationships \newline ~\newline
        PDF/UA (ISO 14289-1) \newline ~\newline
        Best Practices
    }
    & Visual Flow Alignment 
    & The reading order in the document's tags follows the visual flow of content as it appears on the page. \\
    & & Contextual Logic 
    & The sequence is contextually logical, aligning with the academic or narrative structure of the document. \\
    & & Element Positioning 
    & Elements like text, tables, and figures are positioned logically within the reading order; multi-page documents transition smoothly between pages without abrupt breaks. \\
    & & Hierarchy Representation 
    & The reading order conveys the content's hierarchy and organization with no sections of text missing. \\
    & & No Disruptions 
    & There are no overlapping or redundant elements that disrupt logical reading flow. \\
    & & Multi-column Handling 
    & Multi-column layouts are read in the correct logical sequence (not read across columns). \\
    & & Inline Elements 
    & Inline elements (like footnotes, references) are correctly placed in the reading order, not skipped or read out of context. \\
    & & Hidden Content 
    & Hidden or visually obscured content is appropriately skipped in the reading order. \\
    \bottomrule
\end{tabular} 
\caption{Logical reading order and sub-criteria with associated definitions}
\label{tab:reading_order_criteria}
\end{table*}


\begin{table*}[!p]
\small
\renewcommand{\arraystretch}{1.3} 
\begin{tabular}{L{20mm}L{30mm}L{24mm}L{64mm}}
    \toprule
    \textbf{Criterion} & \textbf{Standards} & \textbf{Sub-criterion} & \textbf{Definition} \\
    \midrule
    \multirow[t]{6}{=}{\textbf{Table structure}} 
    & \multirow[t]{6}{=}{%
        WCAG 2.2 SC 1.3.1 Info and Relationships \newline ~\newline
        WCAG 2.2 SC 1.3.2 Meaningful Sequence \newline ~\newline
        WCAG 2.2 SC 2.4.3 Focus Order \newline ~\newline
        WCAG 2.2 SC 2.1.1 Keyboard Accessible \newline ~\newline
        WCAG 2.2 SC 1.3.3 Sensory Characteristics \newline ~\newline
        WCAG 2.2 SC 2.4.6 Headings and Labels \newline ~\newline
        PDF/UA (ISO 14289-1) \newline ~\newline
        Best Practices \\
    }
    & Proper Table Tagging 
    & Data tables are tagged with <Table> elements and contain valid <TR>, <TH>, and <TD> elements in the proper hierarchy. \newline Cells spanning multiple rows/columns (using rowspan/colspan) are tagged semantically. \\
    
    & & Header Semantics
    & Header cells are properly tagged with <TH> elements instead of <TD> elements. \newline Scope attributes (scope="row", scope="col") or headers/id relationships are correctly implemented to associate data cells with their respective headers. \newline For complex tables with grouped or nested headers, header hierarchy is clearly defined. \\

    & & Contextual Clarity 
    & Headers and data labels are contextually clear and unambiguous, ensuring users can understand the table's purpose without relying on visual cues. \\
    
    & & Appropriate Usage 
    & Tables are used appropriately for data presentation rather than for layout purposes; layout tables are marked appropriately if used. \newline No empty or visually hidden table structures are present that serve no semantic purpose.  \\
    
    & & Keyboard Navigability 
    & Tables are navigable with keyboard controls and accessible via assistive technologies without relying on visual cues. \\
    
    & & Complex Table Handling 
    & Nested or split tables are handled correctly; for complex tables, multi-level headers are clearly defined with appropriate programmatic associations. \\
    \bottomrule
\end{tabular} 
\caption{Table structure and sub-criteria with associated definitions}
\label{tab:table_criteria}
\end{table*}


\begin{table*}[!p]
\small
\renewcommand{\arraystretch}{1.3} 
\begin{tabular}{L{20mm}L{30mm}L{24mm}L{64mm}}
    \toprule
    \textbf{Criterion} & \textbf{Standards} & \textbf{Sub-criterion} & \textbf{Definition} \\
    \midrule
    \multirow[t]{5}{=}{\textbf{Functional \newline hyperlinks}} 
    & \multirow[t]{5}{=}{%
        WCAG 2.2 SC 2.4.4 Link Purpose (In context) \newline ~\newline
        WCAG 2.2 SC 2.4.9 Link Purpose (Link only) \newline ~\newline
        WCAG 2.2 SC 1.3.1 Info and Relationships \newline ~\newline
        WCAG 2.2 SC 4.1.2 Name, Role, Value \newline ~\newline
        PDF/UA (ISO 14289-1) \newline ~\newline
        Best Practices
    }
    & Proper Link Tagging 
    & All links are properly tagged as link elements. \newline Cross-references (Internal links) within the document (“see Section 5”) are coded as clickable links rather than just text. \\

    & & Link Functionality 
    & Internal links correctly navigate to the intended reference, page, or section within the document. \newline External hyperlinks use proper link annotations and lead to their URL destinations. \newline All external link targets are valid (no broken links) and named destinations exist for internal links. \\

    & & Link Descriptiveness 
    & External link text provides sufficient context about the destination (e.g., "See research at Nature.com" instead of "Click here"). Exception: Internal reference links need not be verbose \\

    & & Structure Association 
    & Link annotations and <Link> structure elements are correctly associated in the PDF tag structure. \\
    
    & & Link Perceivability 
    & Links are perceivable and discoverable, being accessible via screen reader. \\
    \bottomrule
\end{tabular} 
\caption{Functional hyperlinks and sub-criteria with associated definitions}
\label{tab:links_criteria}
\end{table*}


\begin{table*}[!p]
\small
\renewcommand{\arraystretch}{1.3} 
\begin{tabular}{L{20mm}L{30mm}L{24mm}L{64mm}}
    \toprule
    \textbf{Criterion} & \textbf{Standards} & \textbf{Sub-criterion} & \textbf{Definition} \\
    \midrule
    \multirow[t]{5}{*}{\textbf{Color contrast}} 
    & \multirow[t]{5}{=}{%
        WCAG 2.2 SC 1.4.3 Contrast (Minimum) \newline ~\newline
        WCAG 2.2 SC 1.4.6 Contrast (Enhanced) \newline ~\newline
        WCAG 2.2 SC 1.4.11 Non-text Contrast \newline ~\newline
        Best Practices
    }
    & Small Text Contrast 
    & Text smaller than 18pt (14pt bold) meets the AA-level WCAG contrast ratio of at least 4.5:1 between text and background; AAA-level requires a contrast ratio of at least 7:1. \\
    & & Large Text Contrast 
    & Text 18pt or larger (14pt bold or larger) meets the AA-level WCAG contrast ratio of at least 3:1 between text and background; AAA-level requires a contrast ratio of at least 4.5:1. \\
    & & Informative Visual Elements
    & Essential labels, grid lines, and markers necessary for comprehension meet a minimum contrast ratio of 3:1 against adjacent colors. \newline Color is not the only means of conveying information. \\
    & & Decorative Elements 
    & Decorative or non-essential text is exempted from contrast requirements and should be marked as artifacts in the PDF structure. \\
    & & Text in Images 
    & Text embedded within images, logos, and diagrams complies with the same contrast guidelines as regular text unless it is purely decorative. \\
    \bottomrule
\end{tabular} 
\caption{Color contrast and sub-criteria with associated definitions}
\label{tab:color_criteria}
\end{table*}


\begin{table*}[!p]
\small
\renewcommand{\arraystretch}{1.3} 
\begin{tabular}{L{20mm}L{30mm}L{24mm}L{64mm}}
    \toprule
    \textbf{Criterion} & \textbf{Standards} & \textbf{Sub-criterion} & \textbf{Definition} \\
    \midrule
    \multirow[t]{7}{=}{\textbf{Font readability}} 
    & \multirow[t]{7}{=}{%
        WCAG 2.2 SC 1.4.12 Text Spacing \newline ~\newline
        WCAG 2.2 SC 1.4.8 Visual Presentation \newline ~\newline
        WCAG 2.2 SC 1.4.4 Resize Text \newline ~\newline
        WCAG 2.2 SC 3.2.4 Consistent Identification \newline ~\newline
        PDF/UA (ISO 14289-1) \newline ~\newline
        Best Practices
    }
    & Font Embedding 
    & All fonts are embedded in the PDFs. \newline No fonts are missing or substituted. \\
    
    & & Legibility 
    & Fonts are legible, avoiding decorative or highly stylized fonts for body text. \newline Text uses sans-serif fonts (preferred for digital accessibility) or appropriate serif fonts. \\
    
    & & Typography Usage 
    & Text avoids extended use of all capitals, excessive italics, or script styles. \\
    
    & & Font Size 
    & Body text has a comfortable font size, at least 12pt (16px). No text is smaller than 9pt (12px). \\
    
    & & Semantic Encoding
    & All characters are Unicode-compliant rather than using graphical symbols that may not be properly interpreted by screen readers. \newline Special characters, mathematical symbols, and non-Latin text are properly encoded. \\
    
    & & Font Changes 
    & Font changes (typeface, style, size) are used meaningfully to indicate structure or emphasis, not just for decoration. \\
    
    & & Consistency 
    & Fonts remain consistent across pages with no unexpected changes within the same content section; emphasis (bold/italic) is applied consistently without overuse. \\
    \bottomrule
\end{tabular} 
\caption{Font readability and sub-criteria with associated definitions}
\label{tab:font_criteria}
\end{table*}


\section{Document Modification Details}
\label{sec:document_modification_list}
In Table~\ref{tab:document_modifications}, we detail the specific modifications made to create documents representing different accessibility labels (Passed, Failed, Not Present, Cannot Tell) for each of the seven accessibility criteria.
\begin{table*}[!t]
\centering
\small
\caption{Document Modifications for Each Criterion and Label}
\label{tab:document_modifications}
\renewcommand{\arraystretch}{1.3}
\begin{tabular}{L{20mm}L{33mm}L{33mm}L{23mm}L{25mm}}
\toprule
\textbf{Criterion} & \textbf{Passed (P)} & \textbf{Failed (F)} & \textbf{Not Present (NP)} & \textbf{Cannot Tell (CT)} \\
\midrule
\textbf{Alternative Text Quality} & Included semantically descriptive alternative text that conveyed the purpose and content of images, including key data relationships in charts and diagrams & Introduced common alt text problems: overly generic descriptions (e.g., ``Chart'') or missing key data points & Removed all alternative text while keeping images present & Withheld alternative text information during evaluation \\
\midrule
\textbf{Semantic Tagging} & Ensured proper heading hierarchy, correct list structures, and appropriate semantic content tags throughout the document & Introduced specific errors like misnested headings (h1$\rightarrow$h3$\rightarrow$h2), missing list containers, or non-semantic span tags used for content elements & Removed all structural tags, converting the document to completely untagged content & Provided only the visual rendering of the document without tag tree information during evaluation \\
\midrule
\textbf{Logical Reading Order} & Ensured reading order followed visual layout, with proper handling of multi-column text, and logical progression through document sections & Created disrupted reading flow by rearranging reading order to conflict with visual layout or putting footnotes in the middle of content & Not applicable - all documents have some reading order & Withheld reading order information during evaluation \\
\midrule
\textbf{Table Structure} & Created properly structured tables with correct TH tags for headers, appropriate scope attributes, and logical header-data relationships & Introduced errors such as using TD tags for headers, missing scope attributes, or creating tables with incorrect header-data associations & Evaluated on pages without tables in the document  & Provided only visual representation of tables without any structural information during evaluation \\
\midrule
\textbf{Functional Hyperlinks} & Ensured all hyperlinks were properly tagged, functional, and perceivable & Created issues like non-functional links or improperly tagged links & Removed all hyperlinks from the document while preserving the text content & Withheld link metadata and tag structure information during evaluation \\
\midrule
\textbf{Color Contrast} & Ensured all text met WCAG AA contrast requirements (4.5:1 for normal text, 3:1 for large text) against their backgrounds & Created contrast violations by using color combinations below required ratios for text, particularly in textual content & Not applicable - all documents have some color contrast & Withheld color value information during evaluation, providing only visual rendering \\
\midrule
\textbf{Font Readability} & Used properly embedded, Unicode-compliant fonts with appropriate size, style, and spacing for optimal readability & Introduced problems like non-embedded fonts, decorative fonts for body text, or text set below minimum size requirements & Not applicable - all documents use some fonts & Withheld font metadata information during evaluation, providing only visual rendering \\
\bottomrule
\end{tabular}
\end{table*}

For Logical Reading Order, Color Contrast, and Font Readability, we did not create "Not Present" variants as these properties are inherent to all PDF documents—every document has some reading order, uses colors with some contrast ratio, and employs fonts with certain characteristics.

\clearpage

\section{Prompt and Sample Output}
\label{sec:prompt-output}
We used the following zero-shot prompt to evaluate alternative text quality:

{
\begin{prompt}
You are a PDF accessibility specialist. For each page of a scholarly PDF, analyze all images and their alt text against WCAG 2.2 and PDF/UA guidelines. The input includes the page image, extracted images, and their alt texts. If alternative text is completely missing, do not assume or generate hypothetical alt text.

Accessibility Guidelines
WCAG 2.2 Success Criterion 1.1.1 (Level A) requires that all non-text content presented to the user has a text alternative that serves the equivalent purpose. PDF/UA (ISO 14289-1) section 7.1 requires that all non-decorative content has alternative text.

Follow this protocol:
Step 1: Per-Image Analysis  
For every image on the page, answer these questions for the criteria: 
1. Descriptiveness: Does the alt text fully convey the image's purpose/content within the document context?  
2. Complexity Handling: For complex images (such as charts/graphs/diagrams), does the alt text provide an equivalent understanding of data/relationships and key insights?  
3. Decorative Treatment: If decorative, is alt text empty (`""`) AND tagged as artifact?  
4. Clarity: Is the text free of unexplained jargon and redundant visual descriptions?  
5. Conciseness: Is alt text concise and within recommended length limits (typically 125 characters or fewer when possible)?
6. Contextual Appropriateness: Does alt text consider surrounding content and avoid redundancy by not repeating information already presented in adjacent text?

Step 2: Page Categorization  
Assign the strictest applicable category:  
- Passed (P): The document meets all alternative text quality requirements. All images have appropriate alternative text that satisfies the above criteria.
- Failed (F): The document violates one or more alternative text quality requirements. Some images have inadequate, or inappropriate alternative text.
- Not Present (NP): There are no images in the document with alternative text.
- Cannot Tell (CT): The provided information is insufficient to evaluate alternative text quality, i.e. image alt text is not provided.

Step 3: Overall recommendations for the page:
1. Highlight key issues and suggest actionable fixes for improving alt text quality or clarity.
2. Recommend specific text replacements for problematic alt text.
3. Suggest proper handling for decorative images.
4. Recommend steps for resolving recurring issues across multiple images if applicable.

Step 4: Generate JSON Output  
Based on the provided document information and images, evaluate the alternative text quality and provide your assessment in the following JSON format:

```json
{
  "Page_path": "document.pdf#page=3",
  
  "Images": [
    {
      "Image_path": "chart_2023.png",
      "Alt_Text": "Bar graph of annual revenue"
    }
  ],

  "Page_Evaluation": {
  "accessibility_label": "Passed/Failed/Not Present/Cannot Tell",
  "Recommendations": [
      {
        "Issue": "Description of the issue including its location",
        "Remediation": "Recommended fix for the issue"
      }
    ]
  },
  
  "evaluation_summary": "Brief explanation of your evaluation decision"
}
```
\end{prompt}
}
This comprehensive prompt includes accessibility guidelines, detailed evaluation criteria, clear instructions for accessibility label, guidelines for recommendations, and output formatting requirements.   

A sample model output for 'Failed' accessibility label is shown below:
\begin{jsoncode}
{
  "Page_path": "document.pdf#page=1",
  "Images": [
    {
      "Image_path": "chart_2023.png",
      "Alt_Text": "Bar graph of annual revenue"
    }
  ],
  "Page_Evaluation": {
    "accessibility_label": "Failed",
    "Recommendations": [
      {
        "Issue": "Alt text lacks specific data values and trends",
        "Remediation": "Enhance alt text to describe key data points and trends shown in the revenue chart. Add: '2023 revenue: Q1=\$2.4M, Q2=\$3.1M, Q3=\$4.0M (66\% YoY growth).\'"
      }
    ]
  },
    "evaluation_summary": "Atleast one image in document is complex and the alt text fails to provide an equivalent understanding of data and relationships."
}
\end{jsoncode}

\section{Examples of High-Quality and Low-Quality Recommendations}
\label{sec:recommendation-outputs}

\subsection{Example output of High-Quality Recommendation by GPT-4-Turbo for Table Structure criterion}
\begin{jsoncode}
{
  "Page_path": "document.pdf#page=1",
    "Tables": [
    {
      "Image_path": "table_1.png"
    }
  ],
  "Page_Evaluation": {
    "accessibility_label": "Failed",
    "Recommendations": [
      {
        "Issue": "All header cells in table 1 are tagged as <TD> instead of <TH>. Missing scope attributes or header/id relationships.",
        "Remediation": "Use Adobe Acrobat Pro or similar tools to tag the first row as <TH> and set appropriate scope attributes (e.g., scope='col'). Add scope attributes to <TH> elements or use headers/id relationships to associate header cells with data cells. For the leftmost column, set scope='row' for the header cells."
      }
    ]
  }
}
\end{jsoncode}

This recommendation excels by identifying specific tables and their locations, describing distinct issues and providing technically precise remediation steps.

\subsection{Example output of Low-Quality Recommendation by Gemini-1.5 for Table Structure criterion}
\begin{jsoncode}
{
  "Page_path": "document.pdf#page=1",
    "Tables": [
    {
      "Image_path": "table_1.png"
    }
  ],
  "Page_Evaluation": {
    "accessibility_label": "Failed",
    "Recommendations": [
      {
        "Issue": "Lacks proper <TH> tags for header cells",
        "Remediation": "Ensure all tables have proper header cells.  Use TH tags for header cells and ensure they have appropriate
scope attributes."
      }
    ]
  }
}
\end{jsoncode}

This recommendation lacks specificity about which tables have issues, provides suggestions instead of actionable steps to fix the issue, and omits technical details about what aspects of the table structure need remediation.

\end{document}